\documentclass[epj]{svjour}
\usepackage{epsfig}
\usepackage{cite} 
\begin{document}  
\title{Scattering of light exotic atoms in excited states}
\author{ T.S. Jensen\inst{1,2} \and V.E. Markushin\inst{1}}
\institute{
Paul Scherrer Institute, CH-5232 Villigen PSI, Switzerland
\and 
Institut f{\"u}r Theoretische Physik der Universit{\"a}t Z{\"u}rich,
Winterthurerstrasse 190, CH-8057 Z{\"u}rich, Switzerland 
}
\date{Received: date / Revised version: date}
\abstract{
The  differential and total cross sections for the scattering 
of muonic, pionic, kaonic and antiprotonic hydrogen in excited states 
from atomic hydrogen have been calculated for the purpose of
atomic cascade calculations.  
The scattering problem is treated in a fully quantum mechanical framework 
which takes the energy shifts and, in the case of the hadronic atoms, 
the widths of the $ns$ states into account.
The validity of semiclassical approximations is critically examined.  
 % end of PACS codes
}

\maketitle

\newcommand{\be}{\begin{equation}}
\newcommand{\ee}{\end{equation}}
\newcommand{\non}{\nonumber}
\newcommand{\Rbold}{\mbox{\bf R}}
\newcommand{\rbold}{\mbox{\bf r}}
\newcommand{\dd}{\mbox{\rm d}}

%%%%%%%%%%%%%%%%%%%%%%%%%%%%%%%%%%%%%%%%%%%%%%%%%%%%%%%%%%%%%%%%%%%%%%%%%%%
\section{Introduction}

Exotic hydrogen--like atoms are formed in highly excited states,  
when negative particles ($\mu^-, \pi^-, K^-...$) are stopped in hydrogen. 
The deexcitation of exotic atoms proceeds via many intermediate states  
until the ground state is reached or a nuclear reaction takes place.  
Despite a long history of theoretical and experimental studies 
(see \cite{lb,bl,ma1,ma2} and references therein) 
the kinetics of this atomic cascade is not yet fully understood.  
The current generation of experiments with exotic 
hydrogen--like atoms addresses a number of fundamental problems using 
precision spectroscopy methods, the success of which relies crucially 
on a better knowledge of the atomic cascade.  

In the case of the laser spectroscopy of the Lamb shift in muonic hydrogen 
\cite{ta}, the goal is to determine the proton charge radius with an accuracy 
of $10^{-3}$ from the energy splitting between the $2s$ and $2p$ states.  
This will remove the major theoretical obstacle in the precision calculations of 
the hydrogen Lamb shift, thus extending the limits of the most stringent test of QED 
in a bound system.  The feasibility of this experiment depends on the population 
and the lifetime of the metastable $2s$ state of $\mu^-p$,  and a reliable model 
of the cascade kinetics is essential for this issue.   
The experiment on precision spectroscopy of pionic hydrogen \cite{gotta99pip} 
is expected to determine the $\pi N$ scattering length with a precision 
better than 1\% by measuring the nuclear shifts and widths of the K X--ray lines.   
At this level of precision, the Doppler broadening corrections to the line width 
become important, and they must be reliably calculated from a cascade model.  
In the precision spectroscopy of antiprotonic hydrogen~\cite{gotta99pbp}, the Doppler broadening 
of the L X--ray lines must be taken into account when the $2p$ nuclear widths 
are determined from the X--ray line profile.

  The kinetics of atomic cascade is described by the master equation involving  
all significant processes with the exotic atoms (cascade mechanisms).  
The deexcitation mechanisms include radiative, Auger, and Coulomb processes  
where the transition energy between states with different principle quantum 
number $n$ is carried away mainly by photon, electron, and 
the recoiling particles (including the exotic atom itself), correspondingly.        
  While the deexcitation processes are obviously essential for the 
atomic cascade, the role of the collisional processes preserving 
the principal quantum number $n$ is not less important than 
that of the deexcitation.  The Stark transitions 
$nl\to nl'$ ($l'\neq l$),  affect the population of the $nl$ sublevels. 
Together with the elastic scattering $nl\to nl$ they decelerate the exotic atoms 
thus influencing their energy distribution during the cascade.  
In hadronic atoms, the role of the Stark mixing is especially important 
as it results in a strong absorption during the cascade by feeding 
the $ns$ states which have  absorption widths much larger than 
the states with $l>0$.          

  In the literature starting with the paper of Leon and Bethe~\cite{lb}, 
Stark mixing has often been treated in the semiclassical 
straight--line--trajectory approximation \cite{lb,bl,ve,kl,sc,th}.  
Due to the broad use of this relatively simple model it is desirable  
to know its accuracy in comparison with more advanced and realistic 
quantum mechanical calculations.   By introducing phenomenological 
tuning parameters in the Stark mixing rates \cite{bl,ma1,ma2} one is 
able to reproduce the measured X--ray yields and other experimental data. 
However, the ultimate goal of {\it ab initio} cascade calculations 
demands more accurate results for the collisional processes.

  A fully quantum mechanical treatment based on adiabatic potentials was given 
in \cite{pp1,ppstark,ppdif,pptot}.  However, the shifts and widths of the 
$ns$ states which become important in the final part of the cascade were not 
included in this framework.  
  The deceleration and radiative quenching of muonic hydrogen in the metastable 
$2s$ state were studied in a close--coupling framework in \cite{cf}. 
We reexamined the same problem in \cite{jm99} avoiding some of the approximations 
used in \cite{cf}.   As the close--coupling model can be straightforwardly modified 
to include nuclear absorption in hadronic atoms~\cite{jm00}, it is well suited for
describing the collisional processes during the atomic cascade.  

  In this paper, we present a unified treatment of Stark mixing, 
elastic scattering, and, in the case of hadronic atoms, 
nuclear absorption during collisions.  
For the time being, we restrict our calculations to exotic hydrogen--like atoms.    
  The paper is organized as follows. In sect.~\ref{QM} we present the quantum 
mechanical close--coupling framework used for the calculation of  
the scattering of $x^-p$ atoms in excited states from atomic hydrogen.  
The same processes are treated in the semiclassical approximation in sect.~\ref{SC}. 
The results (differential, partial wave and total cross sections) are discussed
in sect.~\ref{res} and summarized in sect.~\ref{con}. 

Unless otherwise stated, atomic units ($\hbar=e=m_e=1$) are used throughout 
this paper. The unit of cross section is $a_0^2=2.8\cdot10^{-17}\;{\rm cm}^2 $, 
where $a_0 = \frac{\hbar^2}{m_e e^2}$ is the electron Bohr radius.

%%%%%%%%%%%%%%%%%%%%%%%%%%%%%%%%%%%%%%%%%%%%%%%%%%%%%%%%%%%%%%%%%%%%%%%%%%%
\section{Close-coupling calculation of the cross sections}
\label{QM}

\begin{figure}
\begin{center}
\mbox{\epsfysize=3.5cm\epsffile{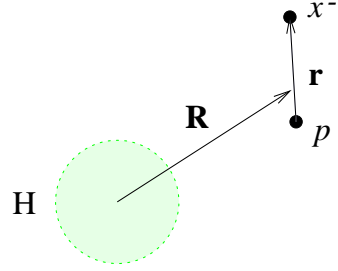}}
\end{center}
\caption{\label{fig:coord}%
Coordinates used for the effective 3--body system 
$x^-p-{\rm H}$: $\Rbold$ is the
vector from the target proton to the center of mass of
the exotic atom, $\rbold$ is the relative vector of the  $x^-p$ system.
The solid angles of $\Rbold$ and $\rbold$ are denoted by 
$\Omega$ and $\omega$, respectively.}
\end{figure}

In this section, the close--coupling framework developed in ref.~\cite{jm99} 
for the 3--body reaction 
\begin{eqnarray}
  (x^-p)_{nl}+{\rm H} & \to & (x^- p)_{nl^\prime}+{\rm H} 
\label{xpH}
\end{eqnarray} 
will be generalized to include absorption effects in hadronic atoms.  
The following notations are used: 
the negative particle $x^-$ with mass $m_{x}$ and the proton with mass $m_p$
form an exotic atom with the total mass $M_{xp}=m_x+m_p$ and 
the reduced mass $\mu_{xp}=m_xm_p/M_{xp}$.  
The coordinates used in the calculations are explained in fig.~\ref{fig:coord}.
The relative orbital angular momentum of $x^-p$ and H is denoted by ${\bf L}$, 
the internal $x^-p$ orbital angular momentum by ${\bf l}$, and 
the total orbital angular momentum by ${\bf J}={\bf L}+{\bf l}$. 

%%%%%%%%%%%%%%%%%%%%%%%%%%%%%%%%%%%%%%%%%%%%%%%%%%%%%%%%%%%%%%%%%%%%%%%%%%%
%%%%%%%%%%%%%%%%%%%%%%%%%%%%%%%%%%%%%%%%%%%%%%%%%%%%%%%%%%%%%%%%%%%%%%%%%%%

%%%%%%%%%%%%%%%%%%%%%%%%%%%%%%%%%%%%%%%%%%%%%%%%%%%%%%%%%%%%%%%%%%%%%%%%%%%
\subsection{The effective $x^-p-{\rm H}$ interaction}

The exotic atom $x^-p$ is described by the Hamiltonian
\be
  H_{xp}=-\frac{\nabla_r^2}{2\mu_{xp}}-\frac{1}{r}+\Delta V
\ee
where $\Delta V$ includes all effects beyond the standard nonrelativistic 
Coulomb two--body problem: vacuum polarization, finite size effects,
and, in case of hadronic atoms, strong interaction between the two particles.  
For the considered systems, $\Delta V$ can be treated as a perturbation 
resulting in the shift $\Delta E_{nl}$ of the Coulomb energy levels: 
\be
 \langle nlm|H_{xp}|nlm\rangle = E_{nl}=-\frac{\mu_{xp}}{2n^2}+\Delta E_{nl}
\ee
where the ket $|nlm\rangle$ denotes the $nlm$ state of the standard Coulomb 
problem.  In the case of hadronic atoms, the atomic cascade is often terminated  
before reaching the ground state due to nuclear reactions like
\begin{eqnarray}
 \pi^-p & \to & \pi^0 n, \ \gamma n\quad ,
\\
  K^-p  & \to & \Sigma^{\pm}\pi^{\mp}, \Sigma^0\pi^0, \Lambda n \quad . 
\end{eqnarray} 
This effect is described by the imaginary part of the complex energy shift
$\Delta E_{nl}$
\be
  {\rm Im}(\Delta E_{nl})=-\Gamma_{nl}/2 
  \label{ImDE}
\ee 
where the width $\Gamma_{nl}$ is the nuclear reaction rate from the $nl$ state. 
For the collisional processes with the exotic hydrogen--like atoms, 
it is sufficient to take into account only the widths of the $ns$ states. 
The $n$ dependence of the hadronic part of the 
complex nuclear shift is described by the 
Deser formula \cite{deser} 
\be
   \Delta E_{ns}^\mathrm{had} = \frac{\Delta E_{1s}^\mathrm{had}}{n^3}\quad . 
\ee  

The Hamiltonian for the $x^-p-{\rm H}$ system is given by
\be
  H = -\frac{\nabla^2}{2\mu} + V(\rbold,\Rbold) + H_{xp}
\ee
where 
\be
 \mu=\frac{M_{xp}M_{\rm H}}{M_{xp}+M_{\rm H}}
\ee
is the reduced mass of the $(x^-p + {\rm H})$ system, 
and $V(\rbold,\Rbold)$ is the Coulomb interaction of the exotic atom with 
the hydrogen electric field.  In most cases of cascade studies, the effects 
of hydrogen excitation or ionization in the collisions without changing $n$ 
can be neglected.     
The electric field of a hydrogen atom in the ground state has the form 
\be
  {\bf F}(\Rbold)=\frac{\Rbold}{R}F(R)
\ee
where 
\be
 F(R)=\frac{1}{R^2}(1+2R+2R^2)e^{-2R}\quad .
\ee
The potential energy is a function of only three variables
\be
  V(\rbold,\Rbold) = V(z^\prime,r^\prime,R)
\ee 
where $\rbold^\prime=(x^\prime,y^\prime,z^\prime)$ is $\rbold$ in
a rotated coordinate system with the $z^\prime$-axis taken along $\Rbold$.  
The electric field of a hydrogen atom is sufficiently strong to mix the $l$ sublevels 
of the $x^-p$ atom during a collision at distances of a few atomic units, $a_0$, 
which are much larger than the size of the $x^-p$ at low $n$.  
Since Stark mixing 
%and the nuclear absorption in collisions are 
is essentially a long--distance process, 
one can use the dipole approximation for the potential
\be
 V(\rbold,\Rbold)=z^\prime F(R)\quad .
\ee 
This can be easily generalized to scattering from other target atoms: one uses
the electric field generated  by the nucleus and the electronic charge density. 

Given the above defined interactions, we solve the time  independent 
 Schr{\"o}dinger equation
\be
 H\psi(\rbold,\Rbold) = E \psi(\rbold,\Rbold)
\ee
with the standard boundary conditions for this multichannel scattering 
problem
% as discussed in Appendix~\ref{variableApp},  
using the close--coupling approximation.   The internal $x^-p$ wave
function is expanded into the set of $n^2$ Coulomb wave functions with the
same quantum number $n$: 
\be
  \psi(\rbold,\Rbold) = R^{-1}\sum_{JMl\Lambda}\phi_{Jl\Lambda}(R)D^{J*}_{\Lambda M}(\Omega )
   \chi_{nl\Lambda}(\rbold^\prime)\quad .
   \label{rotated}
\ee 
The functions $\chi_{nl\Lambda}(\rbold^\prime)$ are the Coulomb wave functions
in a rotating coordinate system with the $z^\prime$--axis chosen as the quantization axis
and $D^J_{\Lambda M}(\Omega )$ are the corresponding rotation functions 
(see appendix~\ref{appd}).  
The functions $\chi_{nl\Lambda}(\rbold^\prime)$ are related to the space fixed functions 
$\chi_{nlm}(\rbold)=r_{nl}(r)Y_{lm}(\omega)$ by 
\be
 \chi_{nlm}(\rbold )=
 \sum_{\Lambda=-l}^{l}D_{m\Lambda}^{l*}(\Omega)\chi_{nl\Lambda}(\rbold ^\prime)\quad .
\ee 
The expansion~(\ref{rotated}) leads to the 
set of $n^2$ coupled second order differential equations for the radial functions
$\phi_{Jl\Lambda}(R)$
\begin{eqnarray}
 && \Big(-\frac{1}{2\mu}\frac{\dd^2}{\dd R^2}+E_{nl}-E\Big) \phi_{Jl\Lambda}(R)
 \nonumber\\
 && +  \sum_{l^\prime \Lambda^\prime}
 \Big( \frac{\langle n; JM\Lambda^\prime l^\prime |{\bf L}^2|n;JM\Lambda l\rangle }{2\mu R^2}\nonumber\\&&+\:
 ^\prime\! \langle nl^\prime  \Lambda^\prime  |V(z^\prime,r^\prime,R)|nl\Lambda\rangle ^\prime\Big)
 \phi_{Jl^\prime \Lambda^\prime }(R)=0\quad .
\label{rad_rotated}
\end{eqnarray}
The basis states $|n;JM\Lambda l\rangle$ are  simultaneous eigenstates of
 $H_{xp}$, ${\bf J}^2$, $J_z$, $J_{z^\prime}$, and ${\bf l}^2$ with
 eigenvalues $E_{nl}$, $J(J+1)$, $M$, $\Lambda$, and $l(l+1)$,
 respectively, and are given by
\begin{eqnarray}
 |n; JM\Lambda l \rangle  & = & 
                        |JM\Lambda\rangle  |nl\Lambda\rangle ^\prime \label{basiswf}\quad ,
\\
 \langle \Omega | JM\Lambda\rangle & = & 
                        \sqrt{\frac{2J+1}{4\pi}}D^{J*}_{M\Lambda}(\Omega )\nonumber\quad ,
\\
 \langle \rbold ^\prime | nl\Lambda \rangle ^\prime & = & 
                           \chi_{nl\Lambda}(\rbold ^\prime) \nonumber\quad .
\end{eqnarray}
The ket $| nl\Lambda \rangle ^\prime$ denotes the eigenstates of the
Coulomb problem in the rotated coordinate system.

Because of rotational invariance the quantum numbers $J$ and $M$ are
conserved and the radial wavefunctions $\phi_{Jl\Lambda}(R)$ are
independent of $M$.
The expansion~(\ref{rotated}) is convenient for computing matrix elements of the 
potential    $V(z^\prime,r^\prime,R)$.  In the dipole approximation, 
the non--vanishing matrix elements of $z^\prime$ correspond to 
$|\Delta l|=1$ and $\Delta\Lambda=0$ where one has
%\MBF{\it primes=?}
\be
^\prime\! \langle n l  \Lambda |z^\prime|n(l-1)\Lambda\rangle ^\prime=
      -\frac{3n}{2\mu_{xp}}\sqrt{\frac{(l^2-\Lambda^2)(n^2-l^2)}{(2l+1)(2l-1)}}\quad .
\ee
The basis states (\ref{basiswf}) are not the eigenstates of ${\bf L}^2$, 
but the matrix elements of ${\bf L}^2$ can be easily obtained by using 
% Following ref.~\cite{ph} one can write ${\bf L}^2$ in the form
\be
 {\bf L}^2=({\bf J}-{\bf l})^2={\bf J}^2+{\bf l}^2-2J_{z^\prime}l_{z^\prime}-
 l_+^\prime J_-^\prime -l_-^\prime J_+^\prime \quad . 
\ee
Together with the results and notations
 of appendix~\ref{appd} this gives  (see also ref.~\cite{ph})
\begin{eqnarray}
 &\langle n;JM\Lambda^\prime l^\prime|&{\bf L}^2|n;JM\Lambda l\rangle =\nonumber\\
 &&\delta_{ll^\prime}\delta_{\Lambda\Lambda^\prime}\Big( J(J+1)+l(l+1)-2\Lambda^2\Big)\nonumber\\
 &&-\delta_{ll^\prime}\Big(\delta_{\Lambda+1\Lambda^\prime}\lambda_-(J,\Lambda)\lambda_+(l,\Lambda)\nonumber\\
 &&+\delta_{\Lambda-1\Lambda^\prime}\lambda_+(J,\Lambda)\lambda_-(l,\Lambda)\Big)\quad .
\label{l2}
\end{eqnarray}
The rotated basis functions were used by Carboni and Fiorentini~\cite{cf} 
to study $(\mu^-p)_{2s}+{\rm H}$ collisions in an approximation where the terms
$ \delta_{ll^\prime}\delta_{\Lambda\Lambda^\prime}(l(l+1)-2\Lambda)$ in 
eq.~(\ref{l2}) were neglected. 

One can get a partial decoupling of the equations (\ref{rad_rotated}) 
by using the following expansion
\be
 \psi (\rbold ,\Rbold )=R^{-1}\sum_{JMLl}\xi_{JLl}(R){\cal Y}_{Ll}^{JM}(\Omega ,\omega )r_{nl}(r)
   \label{coupled}
\ee
where the functions
\be
 {\cal Y}^{JM}_{Ll}(\Omega ,\omega )=
\sum_{M_Lm}\langle LlM_Lm|JM\rangle Y_{LM_L}(\Omega)Y_{lm}(\omega )
\ee
are simultaneous eigenfunctions of 
 ${\bf J}^2$, ${\bf L}^2$, ${\bf l}^2$, and $J_z$ with eigenvalues
 $J(J+1)$, $L(L+1)$, $l(l+1)$, and $M$ respectively. 
The system of the radial Schr{\"o}dinger equations for the functions
$\xi_{JLl}(R)$ has the form
\begin{eqnarray}
 &&\Big(-\frac{1}{2\mu}\frac{\dd^2}{\dd R^2}
 +\frac{L(L+1)}{2\mu R^2}+E_{nl}-E\Big)\xi_{JLl}(R)\nonumber\\
 &&+\sum_{L^\prime l^\prime} \langle  n; L^\prime  l^\prime JM |V(z^\prime,r^\prime,R)|n;LlJM\rangle 
 \xi_{JL^\prime l^\prime }(R)=0
 \label{rad_coupled}\nonumber\\
\end{eqnarray}
with the basis states 
\be
 \langle \Omega,\, \rbold |n;LlJM\rangle = {\cal Y}^{JM}_{Ll}(\Omega,\omega )r_{nl}(r)\quad .
 \label{cbasis}
\ee 
Due to the parity conservation, the value $P=(-1)^{L+l}$ is conserved, and, 
as a result, the $n^2$ differential equations~(\ref{rad_coupled}) are 
decoupled into two sets of $n(n+1)/2$ and $n(n-1)/2$ coupled equations for 
$P=1$ and $P=-1$ correspondingly.   
The systems of the equations~(\ref{rad_rotated}) and
(\ref{rad_coupled}) are related to each other by the linear transformation 
\be
 \phi_{Jl\Lambda}(R)=\sqrt{\frac{2J+1}{4\pi}}\sum_{L}u^{Jl}_{\Lambda L}\xi_{JlL}(R)
\label{xiphi}
\ee
where the coefficients  $u^{Jl}_{\Lambda L}$ are given by eq.~(\ref{uJl}).
The matrix elements of the potential energy  can  then be obtained from
those of the rotated basis by using the coefficients $u^{Jl}_{\Lambda L}$
\begin{eqnarray}
 &\langle n; L^\prime l^\prime JM|&V(z^\prime,r^\prime,R)|n;LlJM\rangle=\nonumber\\
 &&    \sum_{\Lambda\Lambda^\prime}u^{Jl^\prime}_{\Lambda^\prime L^\prime}u^{Jl}_{\Lambda L}\:
^\prime\! \langle nl^\prime  \Lambda^\prime |V(z^\prime,r^\prime,R)|nl\Lambda\rangle ^\prime\quad .
\nonumber\\
\end{eqnarray}

%%%%%%%%%%%%%%%%%%%%%%%%%%%%%%%%%%%%%%%%%%%%%%%%%%%%%%%%%%%%%%%%%%%%%%%%%%
\subsection{Cross sections}

The scattering matrix was calculated numerically using a version of the variable phase 
method~\cite{cal} described in appendix~\ref{variableApp}.  
In order to treat the $ns$ states of hadronic atoms as normal asymptotic states 
the absorptive term (\ref{ImDE}) was switched off for the distances 
between $x^-p$ and ${\rm H}$ larger than $5a_0$. 
The absorption from the $ns$ states between the collisions can be easily taken  
into account by means of a cascade model.      

The use of the dipole approximation in the quantum mechanical framework
makes it necessary to introduce the regularization parameter $R_\mathrm{min}$
as explained in appendix~\ref{variableApp}. The
dependence of  calculated cross sections on $R_\mathrm{min}$ will show how sensitive
the results are to the short distance  behavior (we will show a few examples in
sect.~\ref{res}).

The scattering amplitude for the transition $nlm\to nl^\prime m^\prime$ is given by
\begin{eqnarray}
&f&_{nlm\rightarrow nl^\prime m^\prime }(\Omega)=\frac{4\pi}{2i\sqrt{k^\prime k}}
\sum_{L^\prime L M_{L}^\prime}\Big(i^{L-L^\prime}Y_{L^\prime M_L^\prime}(\Omega )\nonumber\\
&&\times\langle n;L^\prime  l^\prime M_L^\prime m^\prime |S-1|n;Ll0m\rangle Y_{L0}^{*}(0,0)\Big)
\label{Smatr}
\end{eqnarray}
where $\Omega$ is the CMS scattering angle, $k$ and $k^\prime$ are the CMS relative 
momenta of the initial and final state correspondingly.  The S--matrix elements 
in (\ref{Smatr}) are related to the matrix elements between the basis states 
(\ref{basiswf}) by the relation     
\begin{eqnarray}
&&\langle n;L^\prime l^\prime  M_L^\prime m^\prime |S|n;Ll0m\rangle =\nonumber\\
&&\sum_{J}
\langle L^\prime  l^\prime M_L^\prime  m^\prime |Jm\rangle 
  \langle Jm|Ll0m\rangle \langle n;L^\prime l^\prime J m|S|n;LlJm\rangle \quad . \nonumber\\
\end{eqnarray}
The differential and total  cross sections for the transitions 
$nl\to nl^\prime $ are given by
\begin{eqnarray}
\frac{\dd\sigma_{nl\rightarrow nl^\prime}}{\dd\Omega} & = & \frac{1}{(2l+1)}
  \frac{k^\prime}{k} \sum_{m^\prime m}|f_{nlm\rightarrow nl^\prime m^\prime }|^2 \quad , 
\label{diffXS}
\\
  \sigma_{nl\to nl^\prime} & = & \frac{1}{(2l+1)}\frac{\pi}{k^2}\nonumber\\
  &\times&\sum_{JMLL^\prime}|\langle n;L^\prime l^\prime JM|S-1|n;LlJM\rangle |^2  
\nonumber \\ 
   &=& \frac{1}{(2l+1)}\frac{\pi}{k^2}\sum_{J}\Big((2J+1)\nonumber\\
  &\times&\sum_{LL^\prime}|\langle n;L^\prime l^\prime JM|S-1|n;LlJM\rangle |^2\Big)\quad . 
\end{eqnarray}
% for any $M$ ($|M|\leq J$)
The corresponding transport cross sections are given by 
\be
  \sigma^{\rm tr}_{nl\rightarrow nl^\prime }=\int d\Omega (1-\cos\theta)
  \frac{d\sigma_{nl\rightarrow nl^\prime}}{d\Omega} \quad . 
\ee
In the case of hadronic atoms, the scattering matrix is not unitary because of 
the absorption.  The cross sections for the absorption processes are given by
\begin{eqnarray}
 &&\sigma_{nl\to{\rm abs}} = \frac{\pi}{k^2} \sum_{J}\Big((2J+1)\nonumber\\
 &&\times 
  \Big(1 - \frac{1}{(2l+1)} 
       \sum_{LL^\prime l^\prime}|\langle n;L^\prime l^\prime JM|S|n;LlJM\rangle |^2
  \Big)\Big)\quad .\nonumber\\
\label{absXS}
\end{eqnarray} 

  The differential cross sections (\ref{diffXS}) and the absorption cross sections 
(\ref{absXS}) are used in the detailed cascade models as described in \cite{ma1,ma2}.  
When less detailed information is sufficient, $l$--average cross sections defined 
below can be used.  In particular, in those cases where the rates for collisions 
without change in $n$ are much larger than other cascade rates, 
the approximation of the {\it statistically weighted} differential cross section 
is useful: 
\be
 \frac{\dd\sigma_{n-{\rm av}}}{\dd\Omega}=
  \frac{1}{n^2}\sum_{ll^\prime}(2l+1)\frac{\dd\sigma_{nl\to nl^\prime}}{\dd\Omega}
 \label{statd}\quad .
\ee
As a measure of the overall strength of Stark mixing, one can
use the $l$--average Stark cross section: 
\be 
  \sigma_{\rm St}=\frac{1}{n^2}\sum_{l\neq l^\prime}(2l+1)\sigma_{nl\to nl^\prime}
  \quad .  
\label{avstark}
\ee
% which also applies for each value of $J$.

For hadronic atoms with the strong absorption in the $s$ states, we define the 
statistically weighted differential cross section~(\ref{statd}) and the
average Stark cross section~(\ref{avstark}) to include only terms 
with $l>0$ and $l^\prime>0$.
The {\it average absorption} cross section is defined by  
\be
 \sigma_{\rm abs}=\frac{1}{n^2-1}\sum_{l\neq 0}(2l+1)\sigma_{nl\to {\rm abs}}
 \label{avabs}
\ee
which gives a measure of the absorption strength under the assumption of 
statistical population of the $l\neq 0$ sublevels.  Strong absorption
can also take place between the collisions (if the hadronic atom leaves 
the collision zone in an $s$ state)  which is not reflected 
by eq.~(\ref{avabs}).  To this end we define the average cross section 
 \be
 \sigma_{{\rm av}\to ns }=\frac{1}{n^2-1}\sum_{l\neq 0}(2l+1)\sigma_{nl\to ns}
 \label{avS}
\ee
which describes the transitions to the $ns$ state 
from the statistically populated $nl$ sublevels with $l>0$. 
Whether the $s$ state is completely or partially depleted between 
the collisions depends on the type of atom, the quantum number $n$, 
the density of the target, and the kinetic energy.  To estimate the upper limit 
of the absorption, we define the {\it maximum absorption} cross section as the
%\MBF{\it X total}
$l$--averaged sum of the cross sections for nuclear absorption during 
and after collision
\be
 \sigma_{\rm max\; abs}= \sigma_{\rm abs}+\sigma_{{\rm av}\to ns }
\label{maxabsxs}\quad .
\ee

The Stark mixing, deceleration, and absorption rates, which are often used in 
cascade calculations, are defined by the formulas
\begin{eqnarray}
  \lambda_{\rm St}  & = & Nv\sigma_{\rm St}\quad ,\\
  \lambda_{\rm dec} & = & 2\frac{M_{\rm H}M_{xp}}{(M_{\rm H}+M_{xp})^2}Nv\sigma_{\rm tr}\quad ,\\
  \lambda_{\rm abs} & = & Nv\sigma_{\rm abs}  
\end{eqnarray} 
where $N$ is the target density and $v$ the velocity of the exotic atom.
When a significant part of the nuclear reactions takes place
between the collisions, the  absorption is better described by 
the {\it effective absorption rate} defined as following  
%\be 
%  N_{ns}/N_{\rm rest}=
%\frac{\lambda_{{\rm rest}\to ns}}{\Gamma_{ns}+\lambda_{ns\to {\rm rest}}}
%\ee
\be
 \lambda_{\rm eff\; abs} = \lambda_{\rm abs} + 
    \frac{\lambda_{{\rm av}\to ns}}
         {1 + \sum_{l\neq 0}{\lambda_{ns\to nl}}/{\Gamma_{ns}}}\quad .
\label{lambdaeff}
\ee
In the case of very strong absorption during the collisions, the relation   
$\lambda_{{\rm av}\to ns}\ll \lambda_{\rm abs}$ holds, and therefore
\be
  \lambda_{\rm eff\; abs}\approx \lambda_{\rm abs} \qquad
 ({\rm for\;}\Gamma_{ns} \to \infty)\quad .
\ee
When absorption from the $p$ states is important, the effective  rate
for $p$ state absorption is  defined  analogously to   eq.(\ref{lambdaeff}) by considering
statistically populated $l>1$ states.
  A simple comparison of the energy dependent rates for the different
processes cannot substitute detailed cascade calculations using
the detailed cross sections eqs.~(\ref{diffXS},\ref{absXS}) but may be helpful 
for getting a quick overview.  We shall present a few examples in sect.~\ref{res}.

%%%%%%%%%%%%%%%%%%%%%%%%%%%%%%%%%%%%%%%%%%%%%%%%%%%%%%%%%%%%%%%%%%%%%%%%%%%
\section{Semiclassical approximation}
\label{SC}

As the number of coupled second order differential equations 
in the quantum mechanical model of sect.~\ref{QM} grows as $n^2$, 
a simpler framework is desirable for high $n$ states.   
If the collision energy is sufficiently large, one can expect that the relative 
$x^-p-{\rm H}$ motion can be treated classically.  A rough estimate
of the minimum kinetic energy $T$ for which a classical--trajectory description 
is valid can be obtained for the requirement that a large number of partial waves 
$L\sim 2 a_0 k$ ($2a_0$ being the approximate range of the interaction) 
contribute to the cross section; that gives for the kinetic energy of 
muonic hydrogen $T>0.7\;$eV at $L>10$.   
As known from experiment (see \cite{ma2} and references therein),  
the exotic atoms can reach kinetic energies of several eV during the cascade, 
and this makes a semiclassical treatment applicable to many cases of practical 
interest.  The model that has been used most often is
the straight--line--trajectory approximation~\cite{lb,ve,bl,kl,sc,th} where the
small neutral exotic atom is considered as moving along a straight line 
with constant velocity through the electric field of the target 
atom\footnote{Another possibility, which we will not consider
here, is to use deflected trajectories.}.  The time dependent 
electric field  causes transitions among the sublevels of the exotic atom, 
which are treated quantum mechanically.  
This approach was usually used for the calculation of the Stark mixing rates, 
but, as discussed below, differential and absorption cross sections can be 
calculated as well. 

A semiclassical description of our scattering problem is obtained by treating
some of the 6 variables ($\Rbold$ and  \rbold) as classical time dependent variables.  
The remaining variables correspond to the quantum mechanical part of the system 
that is described by the wave function $\psi(t)$ satisfying the Schr{\"o}dinger equation
\be
  i\frac{\partial\psi(t)}{\partial t}=H^\mathrm{SC}(t)\psi(t)  
\ee
where $H^\mathrm{SC}(t)$ depends on $t$ through the classical variables.
The wave function $\psi(t)$ is expanded into a set of orthonormal basis states
\be
 \psi(t)=\sum_j a_j(t)|\alpha_j\rangle
\ee
leading to the time dependent Schr{\"o}dinger equation 
\be
  i\dot{a}_j(t)=\sum_k H^\mathrm{SC}_{jk}(t)a_k(t)
\ee
which must be solved with appropriate boundary conditions. 

To establish a connection to results in the literature we will first discuss 
the simple fixed field model of  Leon and Bethe~\cite{lb,th} where
the $\Rbold$ motion is assumed to be classical and the $x^-p$ atom 
is treated quantum mechanically.  The assumptions are as follows: 
the $x^-p$ moves along a straight line with constant velocity $v$ 
($R(t)=\sqrt{(vt)^2+\rho^2}$ where $\rho$ is the impact parameter), 
only transitions within the $n^2$ states with the given principal 
quantum number $n$ are considered, all the $n^2$ states are degenerate, 
and the electric field from the target atom is directed along the 
quantization axis of the $x^-p$. 
After expanding the internal wave function
of the exotic atom into the Stark eigenstates ($|nn_1\Lambda\rangle$,
$n_1=0,...,n-|\Lambda|-1$)
one is left with a single channel scattering problem
\be
  i\dot{a}_{nn_1\Lambda}(t)=V_{nn_1\Lambda}(R(t))a_{nn_1\Lambda}(t)
\label{idota_lb}
\ee
where 
\be
  V_{nn_1\Lambda}(R)=\frac{3n}{2\mu_{xp}}(2n_1-n+|\Lambda|+1)F(R)\quad .
\ee
Equation~(\ref{idota_lb}) is solved with the boundary condition
\be
 a_{nn_1\Lambda}(-\infty)=1
\ee
 for a range of  values of the impact parameter $\rho$
\be
a_{nn_1\Lambda}(t)=\exp\left( -i \int_{\infty}^{t}V(R(t))\dd t \right)\quad . 
\ee
The eikonal phase shift function~\cite{joa} 
\be
 \chi(\rho)=-\int_{\infty}^{\infty}V(R(t))\dd t 
\ee
is used 
(we take $J+1/2=k\rho$, where $k=\mu v$ is the relative momentum) 
to obtain the scattering amplitude
\be
f^{\rm eikonal}_{nn_1\Lambda}(\theta)=
\frac{1}{2ik}\sum_{J=0}^{J_\mathrm{max}}(2J+1)(e^{i\chi((J+1/2)/k)}-1)
P_{J}(\cos\theta)
\ee
and the  differential cross sections
\be
\frac{\dd\sigma_{nn_1\Lambda}^{\rm eikonal}}{\dd\Omega} = 
           |f^{\rm eikonal}_{nn_1\Lambda}(\theta)|^2\quad .
\ee
This model is not sufficiently accurate for our purposes. 
In sect.~\ref{SM} we generalize it to include different thresholds, 
nuclear absorption during collisions, 
and correct angular coupling between the substates.

%%%%%%%%%%%%%%%%%%%%%%%%%%%%%%%%%%%%%%%%%%%%%%%%%%%%%%%%%%%%%%%%%%%%%%%%
\subsection{Semiclassical model}
\label{SM}

In this case only the radial $R$ motion is considered to be classical, 
the other five variables ($\Omega$, \rbold) are kept quantized.  
The problem of $x^-p$ scattering from hydrogen becomes a 
multichannel scattering problem with different channel momenta and
orbital angular momenta. 
We collect the (complex) energy shifts and the angular part of the 
kinetic energy of the different channels in the diagonal $n^2\times n^2$ matrices 
$\Delta E$ and $L$, respectively.
The collision is specified by the CMS collision energy
\be
 E_\mathrm{cm}=\frac{1}{2}\mu v^2=\frac{k^2}{2\mu}
\ee
and the angular momentum $J$.  
Neglecting the deflection of the neutral $x^-p$ atom 
we take the classical motion to be 
\be
 R=R(t)=\sqrt{(vt)^2+\rho^2}
\ee
where $\rho=\sqrt{J(J+1)}/k$. This introduces some ambiguity into
the model because it requires a common motion $R(t)$ for all
channels. In this paper we  take the common momentum $k$ to be
that of the $l=(n-1)$ states for hadronic atoms while we use
the $ns$ state  in the case of muonic hydrogen.
%(which determine the classical turning point).

We expand the quantum mechanical part of the
system into the states 
\be
 |\alpha_j\rangle=|n;LlJM\rangle
\ee
and obtain the  time dependent Schr{\"o}dinger equation in matrix form  
\be
 i\dot{A}(t)=H^\mathrm{SC}(t)A(t)
 \label{Adot}
\ee
where $H^\mathrm{SC}(t)$ is a $n^2\times n^2$ matrix given by
\be
  H^\mathrm{SC}(t)=\frac{L(L+1)-J(J+1)}{2\mu R^2(t)}+V(R(t))+\Delta E\quad .
\ee
The potential matrix has the elements 
\be
  V_{ij}(R)=\langle n;L^\prime l^\prime JM |V|n;LlJM\rangle
\ee
as in eq.(\ref{rad_coupled}).

The eqs.~(\ref{Adot}) are integrated from $-t_\mathrm{max}$ to $t_\mathrm{max}$, where
$t_\mathrm{max}$ is chosen so large that the potential can be neglected for distances larger than 
$R(t_\mathrm{max})$, with the
boundary conditions 
\be
  A(-t_\mathrm{max})=I\quad .
\ee

The semiclassical scattering matrix is given by
\be
  S^\mathrm{SC}=QA(t_\mathrm{max})Q
\ee
where
$Q$ is the  diagonal matrix given by
\begin{eqnarray}
  Q&=&\exp\Big( i\frac{L(L+1)-J(J+1)}{2\rho k }\arctan (vt_\mathrm{max}/\rho)\nonumber\\
  &&+i\mathrm{Re}(\Delta E) t_\mathrm{max}\Big)\quad .
\end{eqnarray}

With the semiclassical scattering matrix and the formulas from 
sect.~\ref{QM}, results for  differential, total and absorption cross sections
can be obtained.

%%%%%%%%%%%%%%%%%%%%%%%%%%%%%%%%%%%%%%%%%%%%%%%%%%%%%%%%%%%%%%%%%%%%
\subsection{Fixed field model}
\label{FF}

\begin{figure*}
\begin{center}
{\small \hspace{0.1cm} (a) \hspace{6cm} (b) }
\mbox{\epsfysize=5cm\epsffile{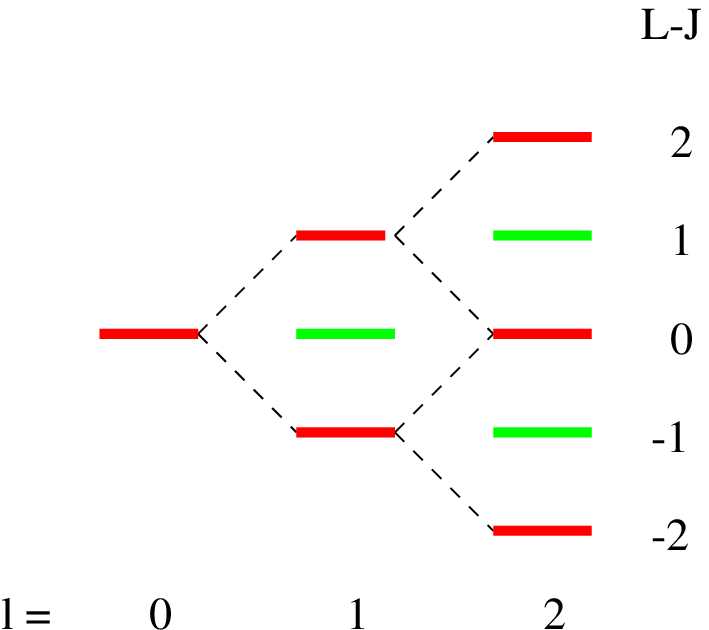}
      \hspace{2cm}
      \epsfysize=5cm\epsffile{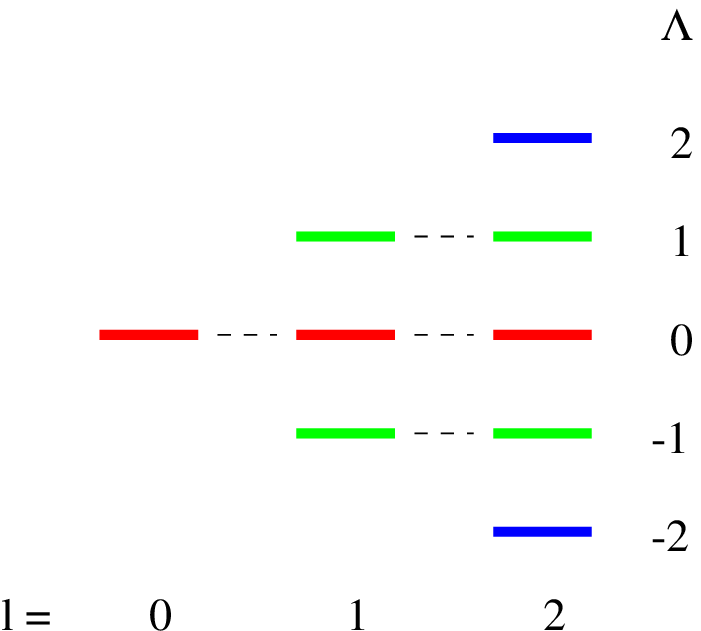}
}
\end{center}
%         %%%\parbox{12cm}
%         %%%{\small \setlength{\baselineskip}{2.6ex} fig.~1. 
\caption{\label{FigLevels}%
Coupling between the sublevels in an $x^-p-{\rm H}$ collision.
The correct coupling (a) conserves parity. In the fixed field model (b)
transitions are only possible between
states with the same eigenvalue $\Lambda$
of $x^-p$ angular momentum along the interatomic axis.   
}
\end{figure*}

The semiclassical approximation simplifies the numerical calculations: 
instead of a system of second order differential equations the same number of 
first order differential equations must be solved.  
A further simplification can be made by neglecting the coupling between 
the internal $x^-p$ angular momentum ${\bf l}$ and the orbital angular 
momentum ${\bf L}$. This corresponds to the approximation  
\be
 L=J
\ee 
in the case discussed above in sect.~\ref{SM}.  In this approximation, 
the quantum number $\Lambda$ 
is conserved in addition to $J$ and $M$ (see fig.~\ref{FigLevels}). 

%%%%%%%%%%%%%%%%%%%%%%%%%%%%%%%%%%%%%%%%%%%%%%%%%%%%%%%%%%%%%%%%%%%%%%%%%%%
%%%%%%%%%%%%%%%%%%%%%%%%%%%%%%%%%%%%%%%%%%%%%%%%%%%%%%%%%%%%%%%%%%%%%%%%%%%

By expanding the solution into the rotated basis states 
($|\alpha_j\rangle=|n;JM\Lambda l\rangle$) one finds the 
time dependent Schr{\"o}dinger equation in matrix form 
\begin{eqnarray}
 i\dot{A}(t)&=&
\left( Z F(R(t))+\Delta E   \right)A(t)\quad ,
\label{idotaff}\\
 A(-t_{max})&=&I
\end{eqnarray}
where
\be
  Z_{ij}=\:
  ^\prime\!\langle nl\Lambda|z^\prime|nl^\prime\Lambda\rangle^\prime\quad .
\ee
Equation~(\ref{idotaff})  must be solved for each value of $\Lambda$, $|\Lambda|<n$. 
If the states 
with  $l>0$ are taken to be degenerate, eq.~(\ref{idotaff}) with $\Lambda\neq 0$
decouple completely in parabolic coordinates~\cite{lb,th} and is easily integrated
as shown above.
This is also the case for $\Lambda=0$ when the $ns$ energy shift is negligible. 
If not, one must solve  $n$ coupled first order equations.  

The fixed field scattering matrix is defined by
\be
S^\mathrm{FF}=Q^\mathrm{FF}A(t_\mathrm{max})Q^\mathrm{FF}
\ee
where the diagonal matrix $Q^\mathrm{FF}$ is given by
\be
 Q^\mathrm{FF}=\exp 
\Big( i\mathrm{Re}(\Delta E) t_\mathrm{max}\Big)\quad .
\ee
The fixed field
scattering amplitude is given by
\begin{eqnarray}
  &&f_{nl\Lambda\to nl^\prime \Lambda}(\theta)=\frac{1}{2ik}\nonumber\\
     &&\times\sum_{J}(2J+1)
     \langle n;JM\Lambda l^\prime|S^\mathrm{FF}-1|n;JM\Lambda l\rangle P_{J}(\cos\theta)\nonumber\\
  \label{amplitudeff}
\end{eqnarray}
and the differential cross section by
\be
  \frac{\dd\sigma_{nl\to nl^\prime}}{\dd\Omega}
  =\frac{1}{2l+1}\sum_{\Lambda}|f_{nl\Lambda\to nl^\prime \Lambda}(\theta)|^2\quad .
\ee
The matrix elements in the r.h.s of eq.(\ref{amplitudeff}) are actually independent
of $M$  since this  quantum number is conserved. 
The cross sections for the processes $nl\to nl^\prime$ and $nl\to {\rm absorption}$ are given
by
\begin{eqnarray}
 &&\sigma_{nl\to nl^\prime}=\frac{1}{2l+1}\frac{\pi}{k^2}\nonumber\\
 &&\times\sum_{J}(2J+1)\sum_{\Lambda}
|\langle n;JM \Lambda l^\prime|S^\mathrm{FF}-1|n;JM\Lambda l\rangle|^2\nonumber\\
\end{eqnarray}
and
\begin{eqnarray}
 \sigma_{nl\to {\rm abs}}&=&\frac{1}{2l+1}\frac{\pi}{k^2} \sum_{J}(2J+1)\Big( (2l+1)  \nonumber\\
 &-&\sum_{\Lambda l^\prime}|\langle n;JM \Lambda l^\prime|S^\mathrm{FF}
|n;JM\Lambda l\rangle|^2\Big)\quad .
\end{eqnarray}

%%%%%%%%%%%%%%%%%%%%%%%%%%%%%%%%%%%%%%%%%%%%%%%%%%%%%%%
%%%%%%%%%%%%%%%%%%%%%%%%%%%%%%%%%%%%%%%%%%%%%%%%%%%%%%%
\section{Results}
\label{res}

Using the methods described in sects.~\ref{QM} and \ref{SC} we have 
calculated the cross sections for the collisions of the   
$\mu^-p$, $\pi^-p$, $K^-p$, and $\bar{p}p$ atoms in excited states with 
hydrogen atoms.   Our calculations had two major goals: first, 
to provide comprehensive sets of the collisional cross sections, which 
are necessary for detailed cascade calculations, and, second, to investigate 
the range of validity of the approximate methods based on the 
semiclassical model and often used in the literature.   
The numerical calculations have been done for the principal 
quantum numbers $n$ and the atomic kinetic energies that are of interest 
for the cascade calculations.   The quantum mechanical framework of 
sect.~\ref{QM} was used for the lower excited states $n=2-5$, 
and the semiclassical calculations were done for the  range of 
$n$ up to $n\sim 10$.  

As the number of the calculated differential cross sections is quite 
large (about 1200 for $\mu^-p$) 
only a small part of the 
results can be shown here, as we describe some main features of the calculated 
cross sections illustrating them with particular examples for different exotic 
atoms.   Concerning the detailed results, they have all been used as 
input for the Monte Carlo kinetics code \cite{mj01,jm01_2}, and the results 
of the cascade 
calculations will be published elsewhere.

%%%%%%%%%%%%%%%%%%%%%%%%%%%%%%%%%%%%%%%%%%%%%%%%%%%%%%%
\subsection{Muonic hydrogen}

\begin{figure}
\epsfig{file=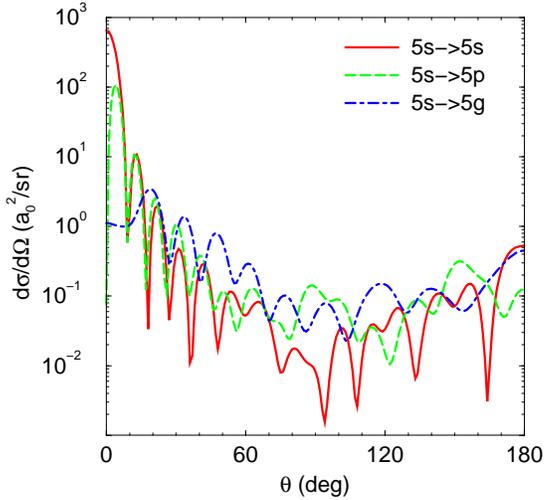, width=7.1cm}
\caption{
The differential cross sections for 
$(\mu^-p)_{5s}+{\rm H}\to(\mu^-p)_{5s,p,g}+{\rm H} $  
vs. CMS scattering angle $\theta$ 
at the laboratory kinetic energy $T=3\;$eV.  
}
\label{fig:dmup}
\end{figure}

The muonic hydrogen scattering is the least complicated case because 
there is no nuclear absorption in the interaction.   The differential 
cross sections are known to have a characteristic shape with a strong 
forward peak and a pattern of maxima and minima \cite{ppdif,jm99,jm00} 
as expected for the interaction that is essentially of a dipole--like type.   
Figure~\ref{fig:dmup} shows an example of the differential cross sections 
for the elastic scattering $5s\to 5s$ and Stark transitions 
$5s\to 5p$ and $5s\to 5g$.   The elastic cross section has a strong peak 
at zero scattering angle, while the Stark transitions reach their maxima
at finite scattering angle.  The peaking in the forward hemisphere 
is much less pronounced for larger changes in quantum number $l$. 

%%%%%%%%%%%%%%%%%%%%%%%%%%%%%%%%%%%%%%%%%%%%%

\begin{figure}
\epsfig{file=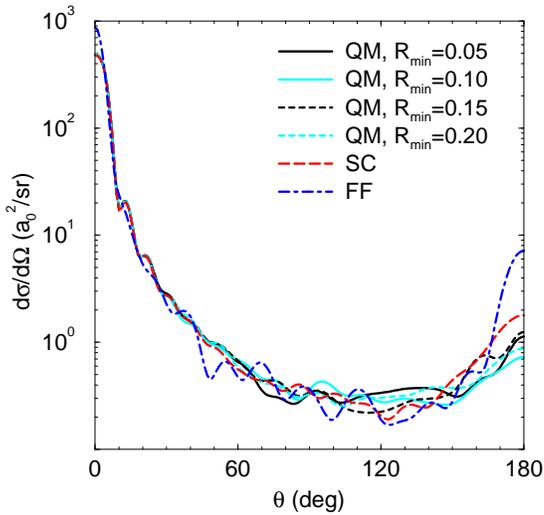, width=7.1cm}
\caption{
The statistically weighted differential cross sections for 
 $(\mu^-p)_{n=5}+{\rm H}\to(\mu^-p)_{n=5}+{\rm H} $
vs. CMS scattering angle $\theta$ 
at the laboratory kinetic energy $T=3\;$eV. 
The fully quantum mechanical (QM)
results computed with $R_{\rm min}$=0.05, 0.1, 0.15 and 0.20
are shown with solid and short--dashed lines, the result of the semiclassical (SC)
model is shown with a dashed line, and that of  
the fixed field model (FF) with a dash--dotted line.
}
\label{fig:dmupav}
\end{figure}
%%%%%%%%%%%%%%%%%%%%%%%%%%%%%%

   To compare the results of different methods from the viewpoint of 
practical applications it is better to look at cross sections averaged over 
some appropriate distribution over $l$ or kinetic energy $T$ 
since many tiny details will be  washed out  anyway in the cascade evolution.   
For the purpose of illustration as well as for simple estimates, 
the statistically weighted cross sections are especially 
useful.   Figure~\ref{fig:dmupav} shows an example of the statistically
weighted differential cross section for $n=5$ calculated in the fully quantum 
mechanical model for four values of the cut--off parameter $R_{\rm min}$
(0.05, 0.1, 0.15 and 0.2)  in comparison with the semiclassical and 
fixed field approximation. 
While the cross sections for the individual transitions $nl\to nl'$ are 
somewhat sensitive to the short range cut off $R_{\rm min}$, the cut--off 
dependence smoothes out in the averaged cross sections.
The semiclassical results are in a good agreement with the quantum mechanical 
ones for energies above 1~eV and $n>2$.    
The fixed field model provides, on average, a fair agreement with the more 
accurate methods for the scattering.
 
\begin{figure}
\epsfig{file=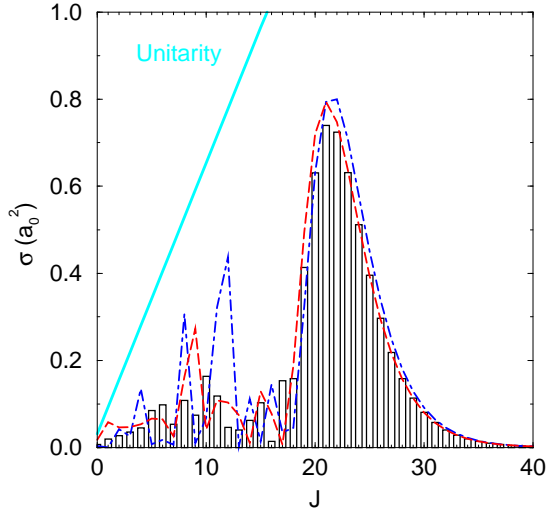, width=7.1cm}
\caption{
Partial wave  cross sections for $(\mu^-p)_{5s}+{\rm H}\to(\mu^-p)_{5p}+{\rm H} $
vs. total angular momentum $J$ 
at the laboratory kinetic energy $T=3\;$eV. 
The fully quantum mechanical results are shown with bars, 
the dashed and dash--dotted lines correspond to the semiclassical and 
the fixed field models. 
}
\label{fig:pmup}
\end{figure}

   Another illuminating way to inspect the complicated structure 
of the differential cross sections is presented in fig.~\ref{fig:pmup} that 
shows the partial wave cross sections for the reaction  
$(\mu^-p)_{5s}+{\rm H}\to(\mu^-p)_{5p}+{\rm H}$ at $T=3$~eV.   
The partial waves can be divided in two groups corresponding to the regimes 
of "weak--coupling" or "strong--coupling" behavior.      
The higher partial waves can be reliably described 
($J \geq ka\approx 18$ for this example) 
in the semiclassical approximation, with the partial cross section showing a smooth 
dependence on the total angular momentum $J$.  

   For the lower partial waves, all $l$ states are strongly mixed with each 
other, and the partial cross sections display a strong dependence on both 
$J$ and $T$.  While the semiclassical approximation is not applicable in 
this situation for individual amplitudes, it still makes a reasonable estimate 
for the average partial cross sections as they are mainly determined by 
the statistical weight of the final states.     
As long as the largest contribution to the total cross section comes from 
the total angular momenta corresponding to the semiclassical regime, 
the semiclassical approximation is adequate for all practical purposes.  
These two regions of $J$ are also different with respect to the dependence 
of the Stark cross sections on the change of the orbital quantum number $l$. 
In the semiclassical regime, the transition amplitude rapidly decreases with 
increasing change in $l$.  In particular, the corresponding partial cross 
sections for the transition $5s\to 5p$ are larger than the ones for the $5s\to 5d$ 
transition.    
  
\begin{figure}
\epsfig{file=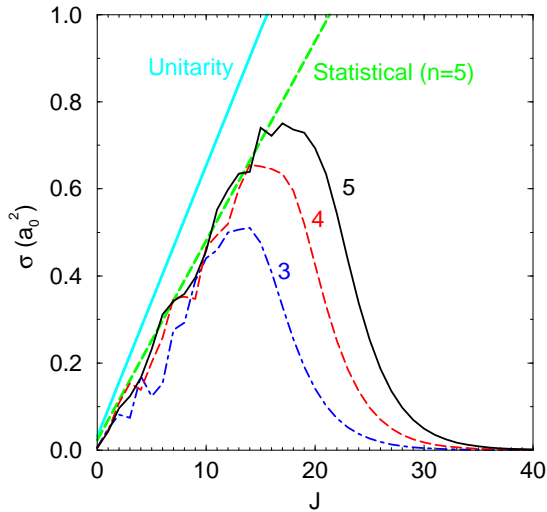, width=7.1cm}
\caption{
Total angular momentum dependence of the average Stark cross sections
for $n=3,4,5$ at the laboratory kinetic energy is $T=3\;$eV. 
The light solid line is the unitarity limit, and the light dashed 
line is the result for the statistical mixing. 
% $n=5$  ?
}
\label{fig:pmupav}
\end{figure}

   Figure~\ref{fig:pmupav} shows the statistical average partial cross sections 
for Stark mixing in the states $n=3,4,5$ in comparison with the unitarity limit.  
In the strong coupling regime, a simple estimate for the average cross section 
can be obtained by assuming that the scattering phases are rapidly changing, 
so that they appear as being "random" (in the old picture of the Stark phase 
accumulated along a trajectory with a small impact parameter it corresponds 
to the so-called complete mixing when the initial state is forgotten after 
the collision).  
For the higher partial waves, the average cross sections are limited not by 
the unitarity constraint, but by the centrifugal barrier which becomes so strong 
that it prevents the exotic atom from getting close to the hydrogen atom.    

%%%%%%%%%%%%%%%%%%%%%%%%%%%%%%%%%%%%%%%%%%%%%%%%%%%%%%%%%%%%%%%%%%%%%%%%%%%%%
%%%%%%%%%%%%%%%%%%%%%%%%%%%%%%%%%%%%%%%%%%%%%%%%%%%%%%%%%%%%%%%%%%%%%%%%%

\begin{figure}
\epsfig{file=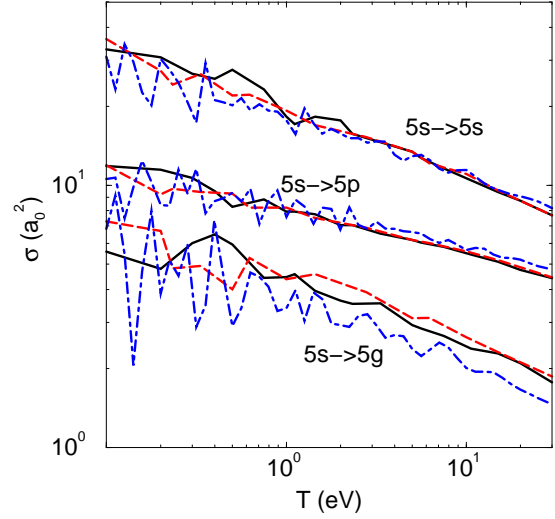, width=7.1cm}
\caption{
Cross sections for $(\mu^-p)_{5s}+{\rm H}\to(\mu^-p)_{5s,p,g}+{\rm H} $ 
vs. laboratory kinetic energy $T$.  The fully quantum mechanical
results (solid lines) are shown in comparison with the results of the 
semiclassical (dashed lines) and the 
the fixed field model (dash--dotted lines).  
}
%\fbox{$T$ or $E_{\mathrm lab}$ ?}
\label{fig:smup}
\end{figure}

The dependence of the total cross sections on the kinetic energy is shown in
fig.~\ref{fig:smup} for the transitions with different change in 
the $\mu^-p$ orbital quantum number $l_f-l_i=0,1,4$:  
$(\mu^-p)_{5s}+{\rm H}\to(\mu^-p)_{5s,p,g}+{\rm H}$. 
As it was said before,  the semiclassical calculations are in a good agreement 
with the quantum mechanical ones for kinetic energies above 1~eV. 
Below 1~eV, where only a few partial waves contribute, the agreement
is still fair after averaging over some energy range.   
The fixed field model is in a fair agreement with the other two.
The cross sections of the fixed field model tend to oscillate more 
than the cross sections computed in the other two models.  For example,  
in the fixed field model the $5s$ state is coupled only to one of the 
$5g$ substates and the corresponding transition is described by one phase shift 
for each partial wave $J$.   
The quantum mechanical and the semiclassical models connect
the $5s$ state with 5 of the 9 substates $5g$ (see fig.~\ref{FigLevels}) 
for a given angular momentum $J$, and the average cross section is smoother as 
it is distributed over a larger number of individual contributions. 

%%%%%%%%%%%%%%%%%%%%%%%%%%%%%%%%%%%%%%%%%%%%%%%%%%%%%%%%%%%%%%%%%%%%%%%%
\begin{figure}
\epsfig{file=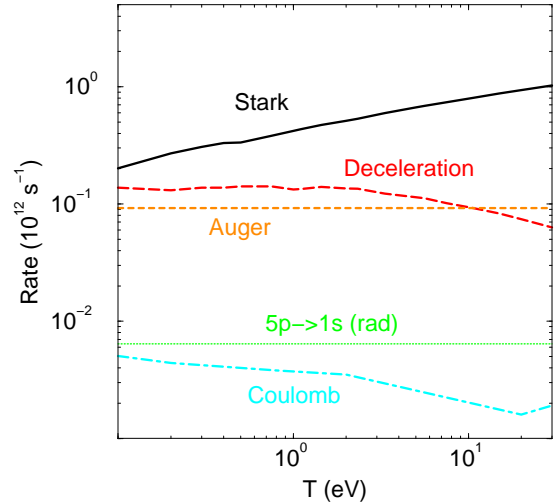, width=7.1cm}
\caption{
The energy dependence of the $l$-average Stark (solid line) and 
deceleration (dashed line) rates for $(\mu^-p)_{n=5}$ at 15~bar  
in comparison with the radiative $5p\to 1s$ rate (dotted line),  
the rates for $n=5\to 4$ external Auger effect~\cite{lb} (light short--dashed line) 
and the $n=5\to 4$ Coulomb deexcitation \cite{psmup} (light dash--dotted line). 
 }
\label{fig:muprates5}
\end{figure}
%%%%%%%%%%%%%%%%%%%%%%%%%%%%%%%%%%%%%%%%%%%%%%%%%%%%%%%%%%%%%%%%%%%%%%%%%

A brief overview of the Stark mixing and the deceleration 
in competition with the deexcitation mechanisms  
is presented in fig.~\ref{fig:muprates5} for a typical example 
of the $\mu^-p$ state $n=5$ in hydrogen gas at 15~bar corresponding to
$0.018$ of liquid hydrogen density.   
The muonic hydrogen atoms arriving at the $n=5$ state during the atomic cascade 
will, on average, undergo a few transitions changing the orbital quantum 
number and loose about half of their kinetic energy before Auger deexcitation 
or, less probably, the radiative transition takes place  
(assuming no other effects, like muonic molecule formation). 
Because the evolution of the kinetic energy distribution is important,  
detailed kinetics cascade calculations~\cite{mj01,jm01_2} are needed to treat  
this problem.

%%%%%%%%%%%%%%%%%%%%%%%%%%%%%%%%%%%%%%%%%%%%%%%%%%%%%%%%%%%%%%%%%%%%%%%%%
\subsection{Pionic hydrogen}

The scattering problem for hadronic atoms is more complicated 
than for the muonic atoms because of the nuclear absorption and 
larger energy shifts.  The previous  fully quantum mechanical calculations based 
on adiabatic potentials~\cite{ppdif,pptot} did not take the shifts and widths 
of the $s$ states into  account, and all the other studies were 
based either on the approximation suggested in \cite{lb} or on the 
time dependent Stark mixing along classical trajectories \cite{rein88,rk}.  
In this section we present the first results of the quantum mechanical 
calculations in comparison with the traditional approximations.   
   The strong interaction shift and width of the $1s$ state of
pionic hydrogen from the final analysis of the PSI experiment 
\cite{schroeder,schr01} were used to calculate the complex energy shifts 
of the $ns$ states:  
%\MBF{update} 
\begin{eqnarray}
   \Delta E_{ns} & = & \frac{\epsilon^{\rm had}_{1s} - i \Gamma_{1s}/2}{n^3} 
                       + \epsilon^{\rm vp}_{ns}\quad ,
\\
   \epsilon^{\rm had}_{1s} & = & -7.11\;{\mathrm eV}\quad ,
\\
   \Gamma_{1s} & = & 0.87\;{\mathrm eV}   
%  \epsilon_{1s}^{\rm had}&=& 7.108\pm 0.013(\rm stat.) \pm0.034(\rm syst.)\;{\rm  eV}\\
%  \Gamma_{1s}&=&0.868\pm 0.040(\rm stat.) \pm 0.038(\rm syst.) \;{\rm  eV}
\end{eqnarray}
where $\epsilon^{\rm vp}_{ns}$ is the  energy shift due to 
the vacuum polarization~\cite{jonsell99}
\begin{eqnarray}
\epsilon^{\rm vp}_{1s}&=&-3.24\;\mathrm{eV}\quad ,\\
\epsilon^{\rm vp}_{2s}&=&-0.37\;\mathrm{eV}\quad ,\\
\epsilon^{\rm vp}_{3s}&=&-0.11\;\mathrm{eV}\quad .
\end{eqnarray}
%\MBF{\it vp = ?} 
The collisions with transitions between the states $l>0$ ($n>2$) are 
qualitatively similar to muonic hydrogen with respect
to Stark mixing and differential cross sections.  
The transitions to the $ns$ states are less probable due to 
the strong interaction energy shift and 
the nuclear reactions taking place during collisions. 

\begin{figure}
\epsfig{file=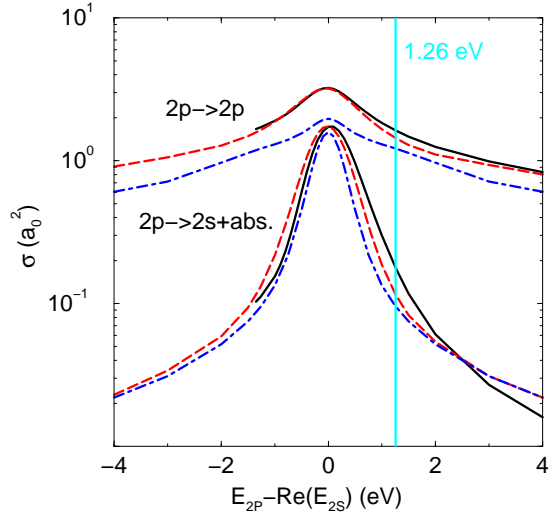, width=7.1cm}
\caption{
The pionic hydrogen cross sections for the elastic scattering $2p\to 2p$ and the  
maximum absorption from the $2p$ state vs. the $2p-2s$ energy difference 
at the laboratory kinetic energy $T=3$~eV.    
The fully quantum mechanical results are shown with solid lines.  
The semiclassical results are shown with dashed lines and those of  
the fixed field model with dash--dotted lines. The light vertical  line
shows the physical value for the energy splitting. The $2s$ width is
$\Gamma_{2s}=0.11$~eV. 
}
\label{fig:pipde}
\end{figure}

To illustrate the effect of the complex energy shift 
we consider the cross sections $\sigma_{2p\to 2p}$ and the maximum absorption
cross section $\sigma_\mathrm{max\; abs}=\sigma_{2p\to {\rm abs}}+\sigma_{2p\to 2s}$ at $T=3\;$eV 
for different unphysical values of the $2p-2s$ energy difference and the widths 
$\Gamma_{2s}$ using the models described in sects.~\ref{QM} and \ref{SC}.    
   Figure~\ref{fig:pipde} shows the dependence on the energy shift for the physical value of 
the width $\Gamma_{2s}=0.11$~eV. 
All three models feature a strong influence of the energy splitting 
$|E_{2p}-\mathrm{Re}(E_{2s})|$ on the Stark mixing with the $2s$ state:   
the nuclear absorption is much more likely when the energy splitting is small 
in comparison with the characteristic Stark splitting in the
electric field of the target atom.  
For small $|E_{2p}-\mathrm{Re}(E_{2s})|$, the semiclassical model agrees well with 
the quantum mechanical results for both cross sections, 
but starts to deviate when the energy splitting is increased.

\begin{figure}
\epsfig{file=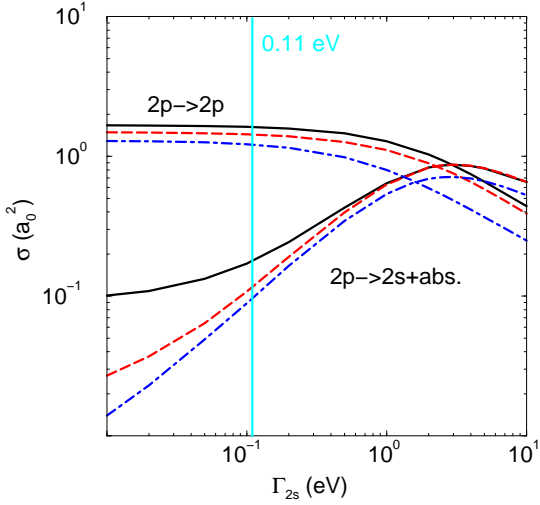, width=7.1cm}
\caption{
The pionic hydrogen cross sections for the elastic scattering
 $2p\to 2p$ and  maximum absorption from the $2p$ state vs. the nuclear
 width $\Gamma_{2s}$ 
at the laboratory kinetic energy $T=3$~eV.       
The fully quantum mechanical results are shown with solid lines.  
The semiclassical results are shown with dashed lines and those of  
the fixed field model with dash--dotted lines. The light vertical line
shows the physical value for the $2s$ width. The $2p-2s$ energy splitting is
1.26~eV. }
\label{fig:pipgam}
\end{figure}

  The dependence of the cross sections on the $\Gamma_{2s}$ with the
energy difference fixed at the physical value 1.26~eV 
(the sum of the strong interaction shift and
the vacuum polarization) is shown  fig.~\ref{fig:pipgam}.   
For $\Gamma_{2s}>0.5\;$eV, the absorption cross sections calculated
in the semiclassical model are in good agreement with the quantum mechanical result, 
while the result of the fixed field model  is about 20~\% lower. 
The good agreement between the semiclassical and the quantum mechanical model 
can be explained as follows: when the nuclear reaction rate is high the
absorption process takes place immediately
from a mixed $2s-2p$ state and is not very sensitive to the kinematics
of the $2s$ channel (the available phase space), which is
treated incorrectly in the semiclassical approximation. 
Both the semiclassical and the fixed field models break down
for smaller $\Gamma_{2s}$, including the physical value, 
 where an accurate treatment of the kinematics is necessary.  

  In general, the fixed field model underestimates the  nuclear absorption
during collision in comparison with the semiclassical model. 
As shown in fig.~\ref{FigLevels}, the fixed field model allows only one of 
the $2p$ substates to be mixed with the $2s$ state and undergo nuclear 
absorption, whereas the correct parity conserving angular coupling used
in our semiclassical model mixes two of the $2p$ substates with
the $2s$ state (in the coupled basis).  For example, for  the physical
values ($E_{2p}-\mathrm{Re}(E_{2s})=1.26$ eV and $\Gamma_{2s}=0.11$~eV) the
maximum absorption cross section at 3~eV is increased by about~18~\% when 
the semiclassical model is used instead of the fixed field model.

%%%%%%%%%%%%%%%%%%%%%%%%%%%%%%%%%%%%%%%%%%%%%%%%%%%%%%%%%%%%%%%%%%%%%%%%%%%%%
%%%%%%%%%%%%%%%%%%%%%%%%%%%%%%%%%%%%%%%%%%%%%%%%%%%%%%%%%%%%%%%%%%%%%%%%%%%%%

%%%%%%%%%%%%%%%%%%%%%%%%%%%%%%%%%%%%%%%%%%%%%%%%%%%%%%%%%%%%%%%%%%%%%%%%%%%%%
\begin{figure}
\epsfig{file=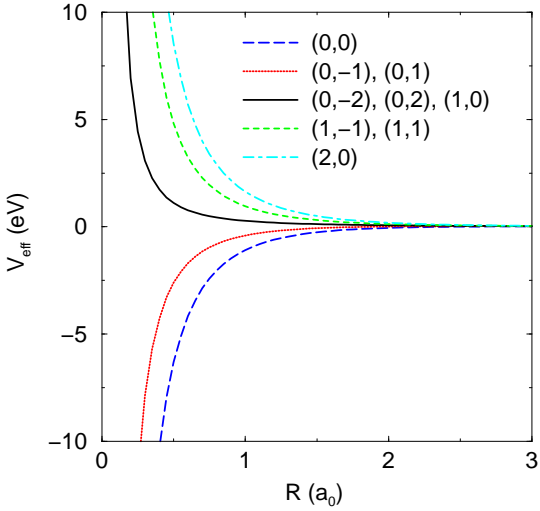, width=7.1cm}
\caption{
The effective potentials for the system
$(\pi^-p)_{n=3}+{\rm H}$ with total angular momentum $J=4$. 
The curves are labeled with the quantum numbers $(n_1,\Lambda)$. 
}
\label{fig:pipj}
\end{figure}
\begin{figure}
\epsfig{file=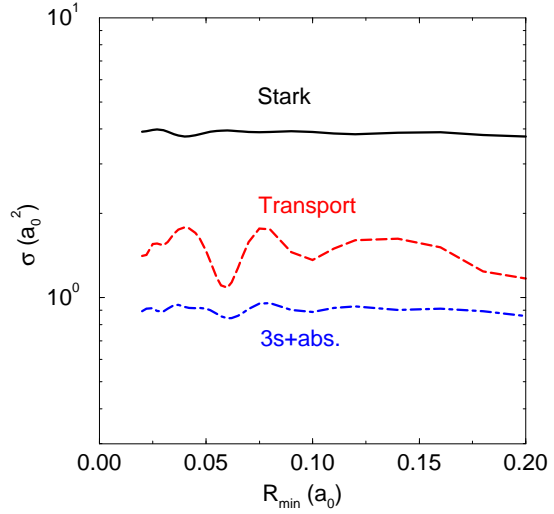, width=7.1cm}
\caption{
$l$--average Stark, transport, and maximum absorption cross sections
for $(\pi^-p)_{n=3}+{\rm H}$ scattering
 vs. short distance cut--off $R_{\rm min}$.
 The laboratory kinetic energy is 3~eV.   
 }
\label{fig:piprmin}
\end{figure}
%%%%%%%%%%%%%%%%%%%%%%%%%%%%%%%%%%%%%%%%%%%%%%%%%%%%%%%%%%%%%%%%%%%%%%%%%

In our quantum mechanical model, the interaction between the exotic atom 
and the target hydrogen atom is approximated by the dipole term, 
which is adequate for large distances.     
When the distance becomes small, this approximation breaks down   
together with the other ones (the close--coupling expansion into the basis  
of atomic states with the same $n$ without taking into account   
symmetry requirements for identical particles).   
Neglecting rotational coupling and energy shifts, the effective
potential energy in the dipole approximation 
for $\pi^-p-{\rm H}$ system is given by
\be
  V_{\rm eff.}(R)=\frac{3n}{2\mu_{\pi^-p}}(2n_1-n+|\Lambda|+1)F(R)+
  \frac{J(J+1)}{2\mu R^2}
 \label{veff}
\ee 
where the parabolic quantum number $n_1$ runs from 0 to $n-|\Lambda|-1$.
Figure~\ref{fig:pipj} shows $V_{\rm eff.}(R)$ as a function
of $R$ for $n=3$ and $J=4$.  Three  of the nine potentials are attractive
for small $R$ and have a $R^{-2}$ singularity in $R=0$.  The
corresponding phase shifts are ill--defined in these cases. When
the correct angular coupling is used and the energy shifts are included,
the potential curves are modified, but the problem with the ill--defined 
phase shifts in the dipole approximation 
remains.  In the present model, the problem is cured
by inserting an infinitely hard sphere with radius
$R_{\rm min}$ around the target nucleus.  For low angular momentum, 
this introduces a dependence of the scattering matrix on 
the cut--off parameter $R_{\rm min}$. 
This should be considered as an uncertainty of the model 
related to the approximate treatment of the short distance behavior.
An example of the $R_{\rm min}$ dependence in the calculation of 
the $(\pi^-p)_{n=3}+H$ scattering is shown in 
fig.~\ref{fig:piprmin} for the $l$--average ($l>0$) Stark, transport, 
and maximum absorption cross sections.  
Both the Stark and the absorption cross sections
are rather insensitive to the value of $R_{\rm min}$, whereas
the transport cross section, which is more sensitive
to the low partial waves, varies moderately ($1.0-1.8\; a_0^2$). However,
this dependence of the transport cross section is less significant when 
the energy distribution in the atomic cascade is taken into account.

\begin{figure}
\epsfig{file=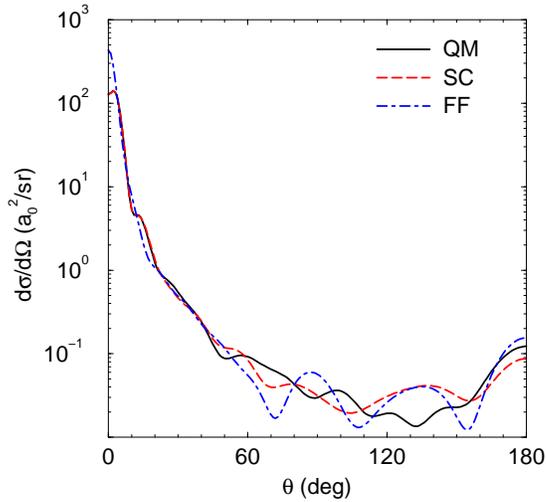, width=7.1cm}
\caption{
Statistically weighted differential cross sections for the
 $(\pi^-p)_{n=3}+{\rm H}$ scattering
vs. CMS scattering angle $\theta$ 
at the laboratory kinetic energy $T=10\;$eV. 
The solid line is the fully quantum mechanical result, 
the dashed line is the semiclassical model,  and 
the dash--dotted line is fixed field model.
}
\label{fig:dpip}
\end{figure}

The statistically weighted differential cross sections for
the $l>0$ sector for $(\pi^-p)_{n=3}+H$
scattering shown in fig.~\ref{fig:dpip} have the similar 
shape as those of  the corresponding
process for $n=5$ in muonic hydrogen (fig.~\ref{fig:dmupav}).
Both the semiclassical and the fixed field model work well 
at kinetic energy $T >1$~eV.
%\MBF{\it energy=?} 
 
%%%%%%%%%%%%%%%%%%%%%%%%%%%%%%%%%%%%%%%%%%%%%%%%%%%%%%%%%%%%%%%%%%%%%%%%%%%%%
\begin{figure}
\epsfig{file=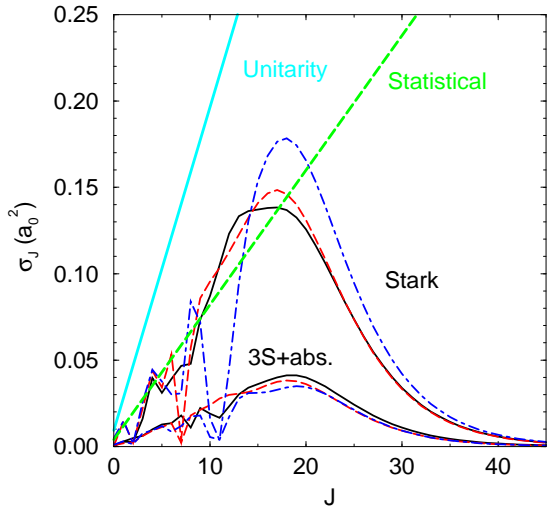, width=7.1cm}
\caption{
The average Stark and maximum absorption  cross sections for the
 $(\pi^-p)_{n=3}+\mathrm{H}$ collisions 
vs. the total angular momentum $J$ 
at the laboratory kinetic energy $T=10\;$eV. 
The solid lines are the fully quantum mechanical results, 
the dashed lines are the semiclassical model,  and 
the dash--dotted lines are the  fixed field model.
The light solid line is the unitarity limit and 
the light dashed line corresponds to  the 
assumption of statistical mixing. 
}
\label{fig:ppip}
\end{figure}
%%%%%%%%%%%%%%%%%%%%%%%%%%%%%%%%%%%%%%%%%%%%%%%%%%%%%%%%%%%%%%%%%%%%%%%%%

The $J$ dependence of the average Stark cross section 
(only the states $l>0$ are included) and the maximum absorption
cross sections (the sum of the absorption during collision and $3p,3d\to 3s$)  
at kinetic energy 10~eV is shown in fig.~\ref{fig:ppip}.    
Like in the case of muonic hydrogen, the contribution to 
the Stark cross section can be divided into two parts: 
a small  $J$ region, where the mixing is strong and the distribution over 
the final states is approximately statistical,  
and a large $J$ region, where the transitions with $|\Delta l|=1$ dominate.  
The semiclassical model is in a good agreement with the quantum mechanical model, 
whereas the agreement with the fixed field model is fair.  All three
models are in a fair agreement for the maximum  absorption cross sections.

\begin{figure}
\epsfig{file=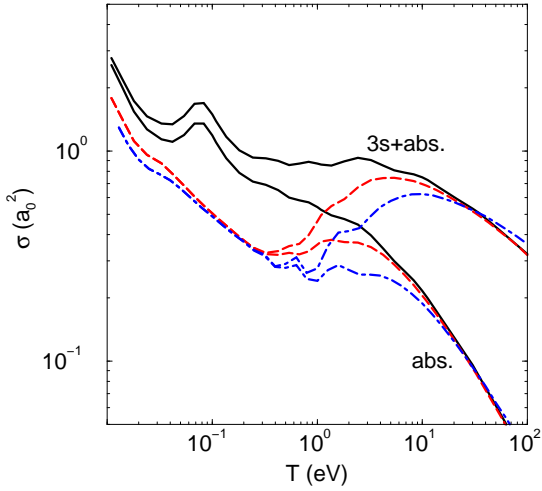, width=7.1cm}
\caption{
The energy dependence of the absorption cross sections, eq.(\ref{avabs}), and 
the maximum absorption cross sections, eq.(\ref{maxabsxs}), 
for pionic hydrogen with $n=3$.   
The solid lines are the fully quantum mechanical results, 
the dashed lines are the semiclassical model,  
and the dash--dotted lines are the fixed field model.  
}
\label{fig:cross3pip}
\end{figure}

The relative role of nuclear absorption during the collisions and 
between the collisions  is illustrated in  fig.~\ref{fig:cross3pip} 
for pionic hydrogen with $n=3$. The absorption cross section 
({\it i.e.} the cross section for nuclear absorption during the
collision, eq.(\ref{avabs})) 
 is shown in comparison with the maximum absorption cross section, eq.(\ref{maxabsxs}).
For energies larger than~$\sim 2$~eV the $\pi^-p$ atom is more likely
to leave the collision zone in the $3s$ state than undergo
nuclear absorption. At high density,  many of the $(\pi^-p)_{3s}$ atoms
with high kinetic energy will leave the $3s$ state before the nuclear reaction
can take place.
The results of the semiclassical and the fixed field model are
in a good agreement with the quantum mechanical results for kinetic 
energies larger than $10$~eV. Below 10~eV the semiclassical description
breaks down and significantly underestimates the absorption cross sections.

\begin{figure}
\epsfig{file=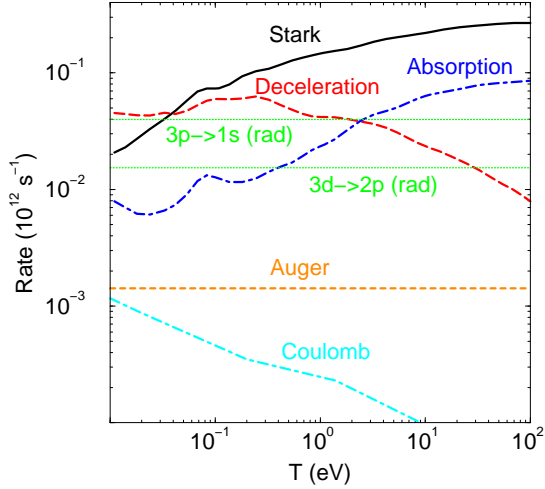, width=7.1cm}
\caption{
The energy dependence of the $l$--average Stark (solid line), 
deceleration (dashed line), and effective absorption (dash--dotted line)  
rates for $(\pi^-p)_{n=3}$ at 15~bar. 
The light dotted lines are the radiative $3p\to 1s$ and $3d\to 2p$ rates. 
The light short--dashed line is the $3\to 2$
Auger deexcitation rate~\cite{lb} and the
light dash--dotted line is the $3\to 2$
adiabatic Coulomb deexcitation rate~\cite{pspip}.
}
\label{fig:piprates}
\end{figure}

Figure~\ref{fig:piprates} shows the rates for different processes for 
the $(\pi^-p)_{n=3}$ state in hydrogen gas at a pressure of 15~bar.  
For energies larger than 0.05~eV Stark mixing is the fastest process 
while the $3p\to 1s$ radiative transition dominates for low energies.  
The effective absorption rate is smaller than the Stark mixing rate   
due to the low statistical weight of the $3s$ state 
and the $3p-3s$ energy difference (only important for low energies).  
The rates for  $3\to 2$ adiabatic Coulomb deexcitation~\cite{pspip}
and external Auger effect~\cite{lb} are also shown for comparison;   
these collisional deexcitation mechanism are obviously suppressed by 
the absorption\footnote{There is experimental evidence for the
$3\to 2$ Coulomb--like deexcitation process in the neutron time-of-flight spectra
in both liquid and gaseous (40~bar) hydrogen~\cite{schott} which can be related
to the formation of excited molecular states, see \cite{jonsell99} and
references therein.}.  
The deceleration rate exceeds the absorption rate below 2~eV, but for higher 
energies the deceleration is suppressed by the absorption, and for energies 
above 20~eV becomes insignificant.

%%%%%%%%%%%%%%%%%%%%%%%%%%%%%%%%%%%%%%%%%%%%%%%%%%%%%%%%%%%%%%%%%%%%%%%%%%%%%
%%%%%%%%%%%%%%%%%%%%%%%%%%%%%%%%%%%%%%%%%%%%%%%%%%%%%%%%%%%%%%%%%%%%%%%%%

%%%%%%%%%%%%%%%%%%%%%%%%%%%%%%%%%%%%%%%%%%%%%%%%%%%%%%%%%%%%%%%%%%%%%%%%%
\subsection{Kaonic hydrogen}

  The nuclear interaction effects in the scattering of the $K^-p$ atoms in 
excited states are even more important than in the case of $\pi^-p$.    
The central values of the KEK result for the $1s$ strong interaction 
shift and width of kaonic hydrogen 
\cite{kek} were used in our calculations: 
\begin{eqnarray}
%   \Delta E_{ns}^\mathrm{had} & = & \frac{(327 - i 203)\;{\mathrm eV}}{n^3}\quad .
  \epsilon_{1s}^{\rm had} & = & 327\pm 63 (\rm stat.)\pm 11 (\rm syst.)\;{\rm  eV}\quad ,
  \\
  \Gamma_{1s} & = & 407\pm 208(\rm stat.)\pm 100(\rm syst.)\;{\rm eV}\quad .
\end{eqnarray}  
% It is expected that both observables will measured with 
% an uncertainty that is a factor ten lower in the DEAR experiment\cite{dear}.
% It is expected that the DEAR experiment\cite{dear} will measure both observables
% with an uncertainty that is a factor ten lower. 
The cascade in kaonic hydrogen differs from that of pionic hydrogen in the initial
condition, it begins 
with a higher $n$ level:
$n_{\rm init}(K^-p)\sim\sqrt{\mu_{K^-p}}\sim 25$ 
as compared to
$n_{\rm init}(\pi^-p)\sim \sqrt{\mu_{\pi^-p}}\sim 15$.
The much larger width of the $ns$ states makes the absorption
during the collisions much more probable than in the $\pi^-p$ atom.
Another important difference between the $K^-p$ and $\pi^-p$ cases is that  
in $K^-p$ the $ns$ energy shift is repulsive, therefore the $nl\to ns$ transitions
are not allowed below the corresponding threshold.  
The strong interaction width of the $2p$ state\footnote{%
It is important for the cascade calculations and strongly influences 
the yield of K X--ray lines, see \protect\cite{kek,mj00}. 
}
is poorly known from the $KN$ scattering data, 
but its effect on the collisional rates is negligible.  

\begin{figure}
\epsfig{file=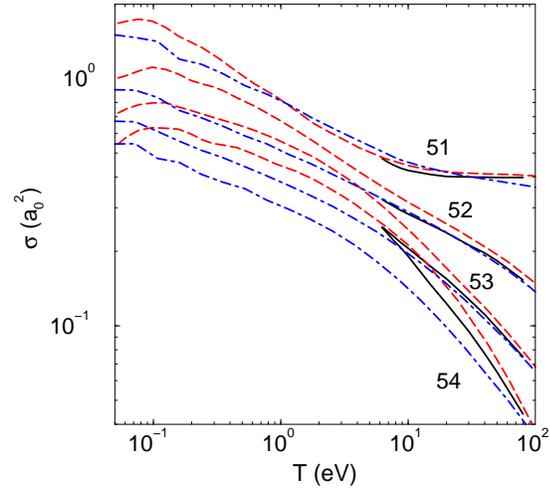, width=7.1cm}
\caption{
Absorption cross sections for $K^-p$ in the $5l$ states with $l=1-4$ scattering
from hydrogen
vs. laboratory kinetic energy.
The fully quantum mechanical results are shown with solid lines,
the semiclassical with dashed lines, and those of the fixed field model 
with dash--dotted lines.
}
\label{fig:Kpcross}
\end{figure}

\begin{figure}
\epsfig{file=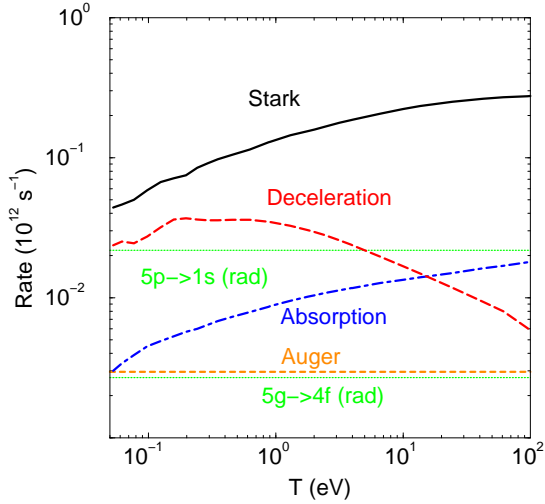, width=7.1cm}
\caption{
The energy dependence of the $l$--average Stark, absorption and deceleration
rates for $(K^-p)_{n=5}$ at 10~bar.  
The results are calculated in the semiclassical model.
The $l$--average Auger rates~\cite{lb} and the radiative rates are shown with 
light lines. 
}
\label{fig:Kprates}
\end{figure}

Figure~\ref{fig:Kpcross} shows an example of the energy dependence of 
absorption cross sections for the states with $n=5$.
The quantum mechanical results are shown only above the $5s$ threshold as 
our numerical algorithm used in this particular case is not reliable in the 
presence of closed channels.  %\MBF{\it discuss}
Overall, there is a fair agreement between the three models above the threshold.  
The fixed field model does, however, result in absorption cross sections that
are somewhat smaller for $l>1$ than those of the semiclassical model.
This can be explained using fig.~\ref{FigLevels}:
the correct angular coupling allows a larger fraction of 
the substates with $l>0$ to be mixed with the $s$ state in a 
single collision and thereby undergo nuclear absorption. 
  Figure~\ref{fig:Kprates} shows the $l$--average Stark, absorption, and 
deceleration rates for $n=5$ calculated in the semiclassical model 
in comparison with the deexcitation rates for a typical gas target at 10~bar.   
The $(K^-p)_{n=5}$ atoms with high kinetic energies are strongly absorbed  
while the radiative and Auger deexcitations dominate in the low energy range.  
Like in the case of pionic hydrogen the situation is complicated
by the deceleration due to elastic collisions.  
The results of cascade calculations based on the presented cross sections 
were discussed in \cite{mj00}.

%%%%%%%%%%%%%%%%%%%%%%%%%%%%%%%%%%%%%%%%%%%%%%%%%%%%%%%%%%%%%%%%%%%%%%%%%%%%%
%%%%%%%%%%%%%%%%%%%%%%%%%%%%%%%%%%%%%%%%%%%%%%%%%%%%%%%%%%%%%%%%%%%%%%%%%

%%%%%%%%%%%%%%%%%%%%%%%%%%%%%%%%%%%%%%%%%%%%%%%%%%%%%%%%%%%%%%%%%%%%%%%%%
\subsection{Antiprotonic hydrogen}

The case of antiprotonic hydrogen is similar to that of $K^-p$: 
the $ns$ nuclear widths are large and the $ns$ nuclear shifts are repulsive. 
The following values \cite{augsburger99} for the spin--averaged
shift and width were used in the present calculations:
%\be
%   \Delta E_{ns}^\mathrm{had}  = \frac{(721 - i 549)\;{\mathrm eV}}{n^3}\quad .
%\ee
\begin{eqnarray}
  \epsilon_{1s}^{\rm had} & = & 721\pm 14\;  {\rm  eV}\quad ,
 \\
  \Gamma_{1s}             & = & 1097\pm 42\;  {\rm  eV}\quad .
%\\  \Gamma_{2p}^{\rm had}   & = & 32.5\pm 2.1\; {\rm  meV} 
\end{eqnarray}  
The hadronic width of the $2p$ state in antiprotonic hydrogen~\cite{augsburger99}
\be
 \Gamma_{2p}^{\rm had}  =  32.5\pm 2.1\; {\rm  meV} 
\ee
is much larger than the radiative one
and absorption during the cascade from the $p$ states is very important. The widths of
the other $np$ states are given by
\be
 \Gamma_{np}^{\rm had}=\frac{32(n^2-1)}{3n^5}\Gamma_{2p}^{\rm had}\quad .
\ee

\begin{figure}
\epsfig{file=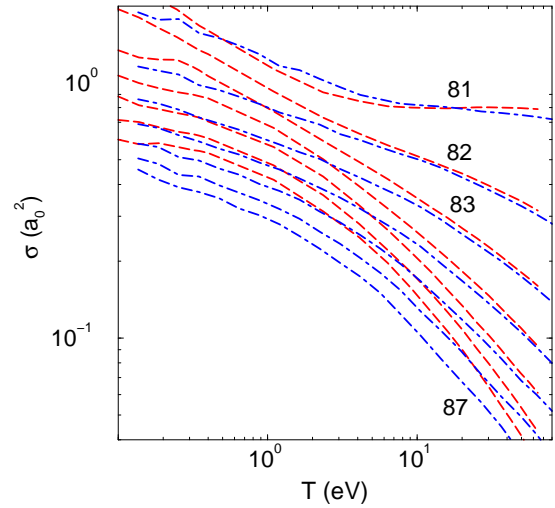, width=7.1cm}
\caption{
Absorption cross sections for $\bar{p}p$ in the $8l$ states with $l=1-7$ 
scattering from hydrogen
vs. laboratory kinetic energy.
The results of the semiclassical model are shown with dashed lines, 
and those of the fixed field model with dash--dotted lines.
}
\label{fig:pbarpabs}
\end{figure}

\begin{figure}
\epsfig{file=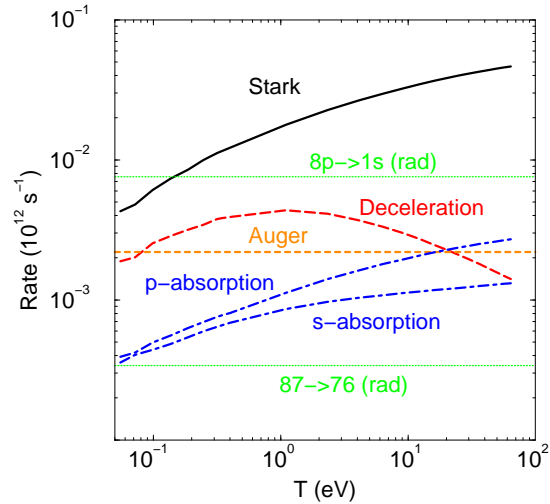, width=7.1cm}
\caption{
The energy dependence of the $l$--average Stark (solid line), 
effective absorption (dash--dotted lines) and deceleration (dashed line)
rates for $\bar{p}p$ for $n=8$  at 1~bar.  
The rates are calculated in the semiclassical model.
The $l$--average Auger rates~\cite{lb} and the radiative rates 
are shown with the light lines. 
}
\label{fig:pbarprates}
\end{figure}

Figure~\ref{fig:pbarpabs} shows the calculated  cross sections for absorption from
the $s$ state at  $n=8$.   
The energy dependence and the $l$ dependence of these cross sections 
are very similar to the kaonic hydrogen case shown in fig.~\ref{fig:Kpcross}. 
The semiclassical and fixed field models are in a fairly good agreement.     
As was observed in the $K^-p$ case, the fixed field model underestimates the
absorption cross sections for $l>1$ due to the approximate treatment
of the angular coupling.
  The rates for the  $l$--average collisional processes for
$(\bar{p}p)_{n=8}$ at a pressure of 1~bar are shown in fig.~\ref{fig:pbarprates} 
in comparison with the deexcitation rates:   
here the absorption just begins, it gets more important at the lower $n$, 
like in the $(K^-p)_{n=5}$ case shown in fig.~\ref{fig:Kprates}, and then eventually 
terminates the cascade. As the $\bar{p}p$ annihilation rate in the $8p$ state is
$10^{12}\;\mathrm{s}^{-1}$, the $p$ state is almost completely depleted between
the collisions at 1~bar. 
The absorption between collisions from the $8p$ state 
is  comparable in strength to the absorption from the $8s$ state
 during collisions for energies below 1~eV
and about two times stronger for higher energies.
The results of detailed cascade calculations for  antiprotonic hydrogen 
will be published elsewhere.

%%%%%%%%%%%%%%%%%%%%%%%%%%%%%%%%%%%%%%%%%%%%%%%%%%%%%%%%%%%%%%%%%%%%%%%%%%%%%
%%%%%%%%%%%%%%%%%%%%%%%%%%%%%%%%%%%%%%%%%%%%%%%%%%%%%%%%%%%%%%%%%%%%%%%%%

%%%%%%%%%%%%%%%%%%%%%%%%%%%%%%%%%%%%%%%%%%%%%%%%%%%%%%%%%%%%%%%%%%%%%%%%%%
\section{Conclusion}
\label{con}

  The total and differential cross sections of Stark mixing and elastic 
scattering have been calculated for the $\mu^-p$, $\pi^-p$, $K^-p$, and 
$\bar{p}p$ atoms for the principal quantum numbers and 
the kinetic energies needed in detailed cascade calculations.  For hadronic 
atoms, the collisional absorption cross sections have been calculated as well.     
For the low states $n=2-5$, the calculations have been done   
in a fully quantum mechanical framework using the close--coupling method.  
For the first time, the effects of nuclear shifts and width of the $ns$ states 
have been taken into account straightforwardly in the quantum mechanical 
scattering problem.  For the intermediate states $n=5-10$, the proposed 
semiclassical model provides an efficient computational method.   
By treating one degree of freedom (the distance between 
$x^-p$ and ${\rm H}$) classically, one can reduce the original system of 
the coupled second order equations to a system of first order equations  
while maintaining the correct angular coupling between the $x^-p$ internal 
angular momentum and the relative orbital angular momentum of 
$x^-p+{\rm H}$ system.  The semiclassical approximation has been found to 
agree fairly well with the fully quantum mechanical calculations, 
%for $n\geq 5$ 
provided the collisional energy is not too low so that 
the number of essential partial waves is large. As the threshold behavior
is not treated correctly in the semiclassical
 approximation, transitions to and from
the $ns$ states cannot be calculated reliably in the near--threshold region.

  Using the above described methods we were able to assess the range of 
validity of the fixed field model, which was commonly used in many cascade 
studies.  
In addition to total Stark mixing and  absorption cross sections calculated 
with this model in the literature~\cite{lb,th}, we have calculated  
the differential cross sections for comparison with 
the more accurate methods. When compared to the semiclassical model,
the fixed field model usually underestimates
 the absorption cross sections
due to the lack of the rotational coupling among the $nl$ sublevels.
When nuclear absorption during collisions is negligible, 
the fixed field model provides, on average, 
a fair description in comparison with the semiclassical approximation for
kinetic  energies larger than a few eV.

The results of this paper have been used in detailed kinetics 
calculations of atomic cascade in $\mu^-p$ and  $\pi^-p$ 
reported in~\cite{mj01}. The detailed description of the results, together
with those in $K^-p$ and $\bar{p}p$, will
be  published in separate papers~\cite{jm01_2}.

  A few problems remain, which are beyond the scope of this paper. 
First, we have considered collisions with {\it atomic} hydrogen. 
One can expect that the molecular structure of the target becomes important 
for large $n$ states when the characteristic size of the exotic atom cannot 
be treated as a small parameter in comparison with the conventional atomic 
scale.  We shall address this problem in a separate paper \cite{jm01_2}.   
Molecular effects~\cite{jonsell99} are also expected to be important at low collisional 
energy when only a small number of molecular ro--vibrational states 
can be excited.  This kinematical region partly overlaps with the 
region of small energies where only a few partial waves are important and 
the dipole approximation is not justified. 
To deal with these problems a genuine many--body framework is needed.

%%%%%%%%%%%%%%%%%%%%%%%%%%%%%%%%%%%%%%%%%%%%%%%%%%%%%%%%%%%%%%%%%%%%%%%%%%%
\section*{Acknowledgment}

We thank P.~Hauser, F.~Kottmann, L.~Simons, D.~Taqqu, and R.~Pohl for 
fruitful and stimulating discussions.
% and M.P.~Locher for useful comments.   

%%%%%%%%%%%%%%%%%%%%%%%%%%%%%%%%%%%%%%%%%%%%%%%%%%%%%%%%%%%%%%%%%%%%%%%%%%
\appendix
\section{The rotation functions $D^J_{\Lambda M}(\alpha,\beta,\gamma )$}
\label{appd}

\renewcommand{\theequation}{A\arabic{equation}}
\setcounter{equation}{0}  
  
This paper uses the conventions of Condon and Shortley for the spherical
harmonics and those of ref.~\cite{vmk} for the 
rotation functions $D^J_{\Lambda M}(\alpha,\beta,\gamma )$. 
Let $\Omega=(\theta,\phi)$ and $\Omega^\prime=(\theta^\prime,\phi^\prime)$ be a 
direction  expressed the original and the rotated coordinates, respectively. The effect
of the rotation on a spherical harmonic is given by the rotation functions   
$D^J_{\Lambda M}(\alpha,\beta,\gamma )$:
\begin{eqnarray}
 Y_{J\Lambda}(\Omega^\prime)&=&
  \sum_{M}D^J_{M\Lambda }(\alpha,\beta,\gamma)Y_{JM}(\Omega )\quad .
% \\
%  Y_{JM}(\Omega)&=&
%  \sum_{\Lambda}D^{J*}_{M\Lambda }(\alpha,\beta,\gamma)Y_{J\Lambda}(\Omega^\prime )
\end{eqnarray} 
 The rotation functions $D^J_{M\Lambda }(\alpha,\beta,\gamma )$ are eigenfunctions 
of the square of the total angular momentum ${\bf J}^2$ and  its projections $J_z$ and 
$J_{z^\prime}$ along
the $z$-axis and the $z^\prime$-axis:
\begin{eqnarray}
   {\bf J}^2 D^J_{M\Lambda }(\alpha,\beta,\gamma)&=&
           J(J+1)D^J_{M\Lambda}(\alpha,\beta,\gamma)\quad ,\nonumber\\
   J_zD^J_{M\Lambda }(\alpha,\beta,\gamma)&=&-M D^J_{M\Lambda }(\alpha,\beta,\gamma)\quad ,\nonumber\\
   J_{z^\prime}D^J_{M\Lambda }(\alpha,\beta,\gamma)&=&
   -\Lambda D^J_{M\Lambda }(\alpha,\beta,\gamma)\quad .
   \label{deq}
\end{eqnarray}
The raising and lowering operators are defined by 
\begin{eqnarray}
 J_{\pm }&=& J_x\pm iJ_y \quad ,\nonumber\\
 J_{\pm }^\prime &=& J_{x^\prime}\pm iJ_{y^\prime}
\end{eqnarray}
and have the properties 
\begin{eqnarray}
 J_{\pm }D^J_{M\Lambda }(\alpha,\beta,\gamma)&=&
 -\lambda_{\mp}(J,M)D^J_{M\mp 1\Lambda }(\alpha,\beta,\gamma)\quad ,\nonumber\\
 J_{\pm }^\prime D^{J}_{M\Lambda }(\alpha,\beta,\gamma)&=&
 -\lambda_{\pm}(J,\Lambda)D^{J}_{M\Lambda \pm 1}(\alpha,\beta,\gamma)
\end{eqnarray}
where 
\be
 \lambda_{\pm}(J,M)=\sqrt{J(J+1)-M(M\pm 1)}\quad .
\ee
In this paper only two of the Euler angles, $\alpha$ and $\beta$, are used.  
Therefore, to simplify notation the rotation functions can be written 
( $\Omega=(\theta,\phi)$):
\be
 D^J_{M\Lambda}(\Omega)=D^J_{M\Lambda}(\phi,\theta,0)\quad .
\ee 
The following integrals represent the normalization and orthogonality conditions 
for the rotation functions and the Wigner--Eckart theorem  for the irreducible 
representations of the rotation group \cite{vmk}    
\begin{eqnarray}    
  &\int d\Omega&\; D^{J*}_{M\Lambda }(\Omega)
   D^{J^\prime}_{M^\prime\Lambda^\prime }(\Omega) =\nonumber\\
   &&\delta_{JJ^\prime}
   \delta_{\Lambda\Lambda^\prime}\delta_{MM^\prime}\frac{4\pi}{2J+1}\quad ,\\
  &\int d\Omega&\; D^{J*}_{MM^\prime}(\Omega)D^L_{M_LM_L^\prime}(\Omega)D^l_{mm^\prime}(\Omega)=
  \nonumber\\
  &&\frac{4\pi}{2J+1}\langle LM_Llm|JM\rangle \langle JM^\prime|LM_L^\prime lm^\prime \rangle
  \label{ddd}
\end{eqnarray}   
where $\langle LM_Llm|JM\rangle$ denotes a Clebsch--Gordan coefficient.
For $\Lambda=0$, the rotation functions are related to the spherical harmonics by 
\be
 D^J_{M0}(\Omega)=\sqrt{\frac{4\pi}{2J+1}}Y^*_{JM}(\Omega)\quad .
 \label{y}
\ee
The coefficients $u^{Jl}_{\Lambda L}$ used in the basis transformation~(\ref{xiphi})
can be found using eqs.~(\ref{ddd}) and
  (\ref{y}) 
%in appendix~\ref{appd}
\begin{eqnarray}
 u^{Jl}_{\Lambda L}&=&\sqrt{\frac{2J+1}{4\pi}}\nonumber\\
 &\times&\int d\Omega d\rbold 
  \Big( D^{J*}_{\Lambda L}(\Omega ) \chi_{nl\Lambda}(\rbold ^\prime)\Big)^*
 {\cal Y}^{JM}_{Ll}(\Omega,\omega)r_{nl}(r)\nonumber\\
 &=&\sum_{M_Lm}\Big(\langle LlM_Lm|JM\rangle \nonumber\\
 &\times& \int  d\Omega\; D^J_{\Lambda M}(\Omega)Y_{LM_L}(\Omega)
  \int d\rbold \chi^*_{nl\Lambda}(\rbold ^\prime)\chi_{nlm}(\rbold )\Big)\nonumber\\
 &=&\sum_{M_Lm}\Big(\langle LlM_Lm|JM\rangle \nonumber\\
 &\times& \int  d\Omega\; D^J_{\Lambda M}(\Omega)Y_{LM_L}(\Omega)D^{l*}_{\Lambda m }(\Omega)\Big)\nonumber\\
 &=&\sqrt{\frac{2L+1}{2J+1}}\langle Ll0\Lambda |J\Lambda\rangle
% =(-1)^{l-\Lambda}\langle Jl-\Lambda \Lambda|L0\rangle 
\label{uJl}
\end{eqnarray}
for any $M$ with $|M|\leq J$. 

%%%%%%%%%%%%%%%%%%%%%%%%%%%%%%%%%%%%%%%%%%%%%%%%%%%%%%%%%%%%%%%%%%%%%%%%%%%%%%%%
\section{Variable phase approach to multichannel scattering}
\label{variableApp}

\renewcommand{\theequation}{B\arabic{equation}}
\setcounter{equation}{0}  
We use a version of the variable 
phase method (see ref.~\cite{cal} and references therein)
to compute the scattering matrix.
In this approach   the problem of solving the second order linear and 
homogeneous
Schr{\"o}dinger equation is transformed into a problem of solving 
a  nonlinear first order equation
for the scattering matrix. 

In this appendix the notation is as follows:
$\xi$ is a column vector containing the radial wave functions, $K$ is the 
diagonal matrix with the channel momenta $K_{mn}=\delta_{mn}k_m$.
The angular momentum quantum numbers for the different channels are 
diagonal elements of the matrix $L$, {\it i.e.} $L_{mn}=\delta_{mn}l_m $.
Then the radial Schr{\"o}dinger equation is given by
\be
 \left( -\frac{\dd^2}{\dd R^2}+\frac{L(L+1)}{R^2}+W(R)-K^2\right)\xi(R) =0
  \label{radw}
\ee 
where $W(R)$ is the reduced potential matrix.

Let $h^{(1)}_l$ and $h^{(2)}_l$ be Riccati--Hankel functions as defined in 
ref.~\cite{cal} and $H_1$ and $H_2$ diagonal matrices with elements
$H_{1mn}(R)=\delta_{mn}h^{(1)}_{l_m}(k_mR)/\sqrt{k_m }$  and
$H_{2mn}(R)=\delta_{mn}h^{(2)}_{l_m}(k_mR)/\sqrt{k_m }$.

The  scattering matrix $S(R_0)$ obtained from  eq.~(\ref{radw}) with $W$ truncated at 
$R_0$ ({\it i.e.} with the substitution 
$W(R)\rightarrow W(R)\theta (R_0-R)$) is a function of 
$R_0$ and satisfies the equation 
\be
S^\prime =\frac{i}{2}(SH_1-H_2)W(H_1S-H_2)\quad .
\label{s}
\ee 
The scattering matrix of the full problem is given by $S=S(\infty )$.

In eq.~(\ref{s}) the dependence on the angular momentum is contained 
in the Riccati--Hankel
functions. In numerical calculations,
 it can  be more convenient to combine the 
potential and the angular momentum term in one effective potential
\be
 W_{eff}(R)=W(R)+\frac{L(L+1)}{R^2}\quad .
\ee
Following the same procedure  as in the derivation of eq. (\ref{s}) one obtains  
\be
\bar{S}^\prime =\frac{i}{2}(\bar{S}\bar{H}_1-\bar{H}_2)W_{eff}(\bar{H}_1\bar{S}-\bar{H}_2)
\label{sbar}
\ee 
where $\bar{H}_1(R)=K^{-1/2}e^{iKR}$ and $\bar{H}_2(R)=K^{-1/2}e^{-iKR}$.  
%where $\bar{H}_1(R)$ and $\bar{H}_2(R)$ are diagonal with elements
%$k_j^{-1/2}\bar{h}_{l_j}^{(1)}(k_jR)=k_j^{-1/2}e^{i}
The matrix $\bar{S}(R)$ is related to the scattering matrix  by
\be
 S=e^{-i\pi L/2}\bar{S}(\infty )e^{-i\pi L/2}\quad .
\ee
For short range potentials  eq.~(\ref{sbar}) appears to be less 
suitable than  eq.~(\ref{s}) because the solution must be obtained 
for large values of $R$ due to the long range behavior of centrifugal repulsion. 
This is, however, not necessary because it is always possible to 
convert $\bar{S}(R)$ to $S(R)$ and vice versa with algebraic methods.

The connection between $S(R)$ and $\bar{S}(R)$ can be established
through the identity of the wave functions and their derivatives
in $R$. Let $\Xi$ be a square matrix with linear independent solutions $\xi$
as columns. $\Xi$ and $\Xi^\prime$ can be expressed both in terms of 
$S$ and $\bar{S}$
\begin{eqnarray}
\Xi&=& (H_1S-H_2)N=(\bar{H}_1\bar{S}-\bar{H}_2)\bar{N}\nonumber\\
\Xi^\prime&=& (H_1^\prime S-H_2^\prime )N=(\bar{H}_1^\prime \bar{S}-\bar{H}_2^\prime )\bar{N}
\label{XiXiprime}
\end{eqnarray}
where $N$ and $\bar{N}$ are square matrices. 
From the relations (\ref{XiXiprime}), one finds the following 
expression for $S$
\be
S=H_1^{-1}\Big( (\bar{H}_1\bar{S}-\bar{H}_2)\bar{N}N^{-1}+H_2 \Big)
\ee
with
\begin{eqnarray}
 \bar{N}N^{-1}&=&\Big( H_1^\prime (\bar{H}_1\bar{S}-\bar{H}_2)-H_1
(\bar{H}_1^\prime \bar{S}-\bar{H}_2^\prime )\Big)^{-1}\nonumber\\
&\times&(H_1H_2^\prime-H_1^\prime H_2)\quad .
\end{eqnarray}

We compute the scattering matrix for  the $x^-p+H\to x^-p+H$ 
process by solving eq.~(\ref{sbar}) in the  coupled basis with
the boundary condition that  $\bar{S}(R_{\rm min})$ is a diagonal
matrix with elements 
\be
\bar{S}_{mn}=\delta_{mn}\frac{1+i\tan (k_m R_{\rm min})}{1-i\tan (k_m R_{\rm min})} 
\ee
which is the scattering matrix for S--wave scattering from an
infinitely hard sphere with radius $R_{\rm min}$.  The effect of nuclear absorption
from the $ns$ states is included by adding the imaginary part (the real part
is already taken into account in the momentum matrix $K$)
of the $ns$ energy shift to the potential
\begin{eqnarray}
W_{ij}(R)&=&2\mu\Big( \langle n;L^\prime l^\prime JM|V(R)|n;LlJM\rangle\nonumber\\
&+&\delta_{l0}\delta_{l^\prime 0} (-i\Gamma_{ns}/2) \theta(R_0-R)    \Big)\quad .
\end{eqnarray}  
The nuclear absorption is turned off for distances larger
than $R_0$; we use  $R_0=5a_0$ in this paper.  This allows the mixing 
$nl\leftrightarrow ns$ 
to take place during the collision together with the absorption effects 
while the $ns$ states remain well defined asymptotic states.        
The nuclear reactions that occur after the collision are taken into 
account by a cascade model.
 
It is necessary to set the boundary condition  away from $R=0$ because the 
potential taken in the dipole approximation has an $R^{-2}$ singularity 
at $R=0$ which makes the Schr{\"o}dinger equation ill--defined.  
The exact potential, however, becomes finite for small $R$. 
By using the short distance cut--off $R_{\rm min}$ we obtain a
well--defined Schr{\"o}dinger equation, but  the calculated
cross sections become  dependent on this regularization parameter.  By varying
$R_{\rm min}$ around the value where the exact potential becomes 
repulsive one can get an estimate of the uncertainty of the cross sections 
due to  the treatment of the short distance behavior. Throughout this paper 
we use the value $R_{\rm min}=0.05a_0$ unless otherwise stated.

%%%%%%%%%%%%%%%%%%%%%%%%%%%%%%%%%%%%%%%%%%%%%%%%%%%%%%%%%%%%%%%%%%%%%%%%%%%
\newcommand{\etal}{\mbox{\it et al.}}

\bibliographystyle{unsrt}

\begin{thebibliography}{99}

\bibitem{lb} 
        M. Leon and H.A. Bethe, Phys. Rev. {\bf 127}, 636 (1962) 
\bibitem{bl} 
        E. Borie and M. Leon, Phys. Rev. A {\bf 21}, 1460  (1980)

\bibitem{ma1} 
        V.E. Markushin, Phys. Rev. A {\bf 50}, 1137 (1994)
\bibitem{ma2} 
        V.E. Markushin, Hyperfine Interaction {\bf 119}, 11 (1999)  

\bibitem{ta} 
        D. Taqqu \etal, Hyperfine Interactions {\bf 119}, 311 (1999) 
% more about mup experiment ?

\bibitem{gotta99pip} 
        D.~Gotta, $\pi N$ Newsletter {\bf 15},  276 (1999)
\bibitem{gotta99pbp}
       D. Gotta \etal, Nucl. Phys. A {\bf 660}, 283 (1999)       

\bibitem{ve} 
        J.-L. Vermeulen, Nucl. Phys. B {\bf 12}, 506 (1969)
\bibitem{kl} 
        G. Kodosky and M. Leon, Nuovo Cimento {\bf 1B}, 41 (1971) 
\bibitem{sc} 
        M.C. Struensee and J.S. Cohen, Phys. Rev. A {\bf 38}, 44 (1988) 
\bibitem{th} 
        T.P. Terada and R.S. Hayano, Phys. Rev. C {\bf 55}, 73 (1997)

\bibitem{pp1}
        V.P. Popov and V.N. Pomerantsev, Hyperfine Interactions {\bf 101/102}, 133 (1996)
\bibitem{ppstark}
        V.P. Popov and V.N. Pomerantsev, Hyperfine Interactions {\bf 119},  133 (1999)
\bibitem{ppdif} 
        V.P. Popov and V.N. Pomerantsev, Hyperfine Interactions {\bf 119}, 137 (1999) 
\bibitem{pptot}
        V.V. Gusev, V.P. Popov and V.N. Pomerantsev, 
        Hyperfine Interactions {\bf 119}, 141  (1999)

\bibitem{cf} 
        G. Carboni and G. Fiorentini, Nuovo Cimento {\bf 39B}, 281 (1977)  

\bibitem{jm99}
        T.S. Jensen and V.E. Markushin, PSI-PR-99-32 (1999), nucl-th/0001009
\bibitem{jm00}
        T.S. Jensen and V.E. Markushin, Nucl. Phys. A {\bf 689}, 537  (2001)
        
\bibitem{deser} 
        S. Deser et al., Phys. Rev.  {\bf 96}, 774  (1954) 
\bibitem{ph} 
        R.T. Pack and J.O. Hirschfelder, J. Chem. Phys. {\bf 49},  4009 (1968)
\bibitem{cal}
        F. Calogero, {\it  Variable Phase Approach to Potential Scattering } 
        (New York: Academic Press, 1967)
\bibitem{joa}
        C.J. Joachain, {\it Quantum Collision Theory} (North--Holland
        Publishing Company, Amsterdam, 1975)
\bibitem{mj01}
       V.E. Markushin and T.S. Jensen, Hyperfine Interactions  (in press)       
\bibitem{jm01_2}
        T.S. Jensen and V.E. Markushin (to be published)
\bibitem{psmup} 
        L.I. Ponomarev and E.A. Solov'ev, Hyperfine Interactions {\bf 119}, 55 (1999) 
\bibitem{rein88}
        G.~Reifenr{\"o}ther et al., Phys. Lett. {\bf 214 B}, 325 (1988)
\bibitem{rk} 
        G. Reifenr{\"o}ther and E. Klempt, Nucl. Phys. A {\bf 503},  885 (1989)
\bibitem{schroeder}
        H.-Ch. Schroeder \etal, Phys. Lett. B {\bf 469}, 25 (1999)      
\bibitem{schr01}
        H.-Ch. Schroeder \etal,
        % Preprint ETHZ-IPP PR-2001-01 (2001)
        Eur. Phys. J. C {\bf 21}, 473 (2001)
\bibitem{jonsell99}
        S.~Jonsell, J.~Wallenius and P.~Froelich, Phys. Rev. A {\bf 59}, 3440 (1999) 
\bibitem{pspip} 
        L.I. Ponomarev and E.A. Solov'ev, JETP Lett. {\bf 64}, 135  (1999) 
\bibitem{schott} 
        J. Schottm{\"u}ller \etal, Hyperfine Interactions  {\bf 119}, 95 (1999)
\bibitem{kek}
        M. Iwasaki \etal, Phys. Rev. Lett. {\bf 78}, 3067 (1997) 
\bibitem{mj00}
        V.E. Markushin and T.S. Jensen, Nucl. Phys. A {\bf 691}, 318  (2001)
\bibitem{augsburger99}
       M. Augsburger \etal, Nucl. Phys. A {\bf 658}, 149 (1999)
\bibitem{vmk} 
        D.A. Varshalovich, A.N. Moskalev, V.K. Khersonskii: 
        {\it Quantum Theory of Angular Momentum} 
        (World Scientific Publishing Co Ptc. Ltd., Singapore, 1988) 



\end{thebibliography}

\end{document}